\documentclass[sigconf]{acmart}

\usepackage{booktabs} 

\usepackage{amsmath}
\usepackage{subfigure}
\usepackage{graphicx}
\usepackage{amssymb}
\usepackage{fixmath}
\usepackage{caption}
\usepackage{algorithmic}
\usepackage{algorithm}

\usepackage{algorithmic}
\usepackage{algorithm}
\usepackage{multirow}

\usepackage{flushend}

\begin{document}
\title{WNOS: An Optimization-based Wireless Network Operating System}



\author{Zhangyu Guan, Lorenzo Bertizzolo, Emrecan Demirors, Tommaso~Melodia}
\affiliation{%
  \institution{Department of Electrical and Computer Engineering\\Northeastern University, MA 02115, USA}
}
\email{[zguan, bertizzolo, edemirors, melodia]@ece.neu.edu}

%
%
%
%
%


\begin{abstract}
Applications of { software defined networking (SDN)} concepts to infrastructure-less wireless networks are substantially unexplored, mainly because of the complex  nature of the distributed
control problems and of the unavailability of a high-speed backhaul. This article presents an initial attempt at \emph{developing a principled approach to software-defined wireless networking based on cross-layer optimization theory, and at bridging the gap between software defined networking and distributed network optimization}.

We investigate the basic design principles for a new \emph{Wireless Network Operating System (WNOS), a radically different approach to SDN for infrastructure-less wireless networks.} Departing from  well-understood approaches inspired by OpenFlow, WNOS provides the network designer with an abstraction hiding (i) the lower-level details of the wireless protocol stack and (ii) the distributed nature of the network operations.
Based on this abstract representation, the WNOS takes network control programs written on a centralized, high-level
view of the network and automatically generates distributed cross-layer control programs based on distributed optimization theory that are executed by each individual node on an
abstract representation of the radio hardware.

We first discuss the main architectural principles of WNOS. Then, we discuss a new approach to generate solution algorithms for each of the resulting subproblems in an automated fashion. Finally, we illustrate a prototype implementation of WNOS on software-defined radio devices and test its effectiveness by considering specific cross-layer control problems. Experimental results indicate that, based on the automatically generated distributed control programs, WNOS achieves 18\%, 56\% and 80.4\% utility gain in networks with low, medium and high levels of interference; maybe more importantly, we illustrate how the global network behavior can be controlled by modifying a few lines of code on a centralized abstraction.
\end{abstract}


%
%
%

\keywords{Software-defined Networking, Distributed Optimization, Automated Control Program Generation, Wireless Ad Hoc Networks.}

\maketitle

\begin{figure*}[t]
\centering
\includegraphics[width=0.8\textwidth]{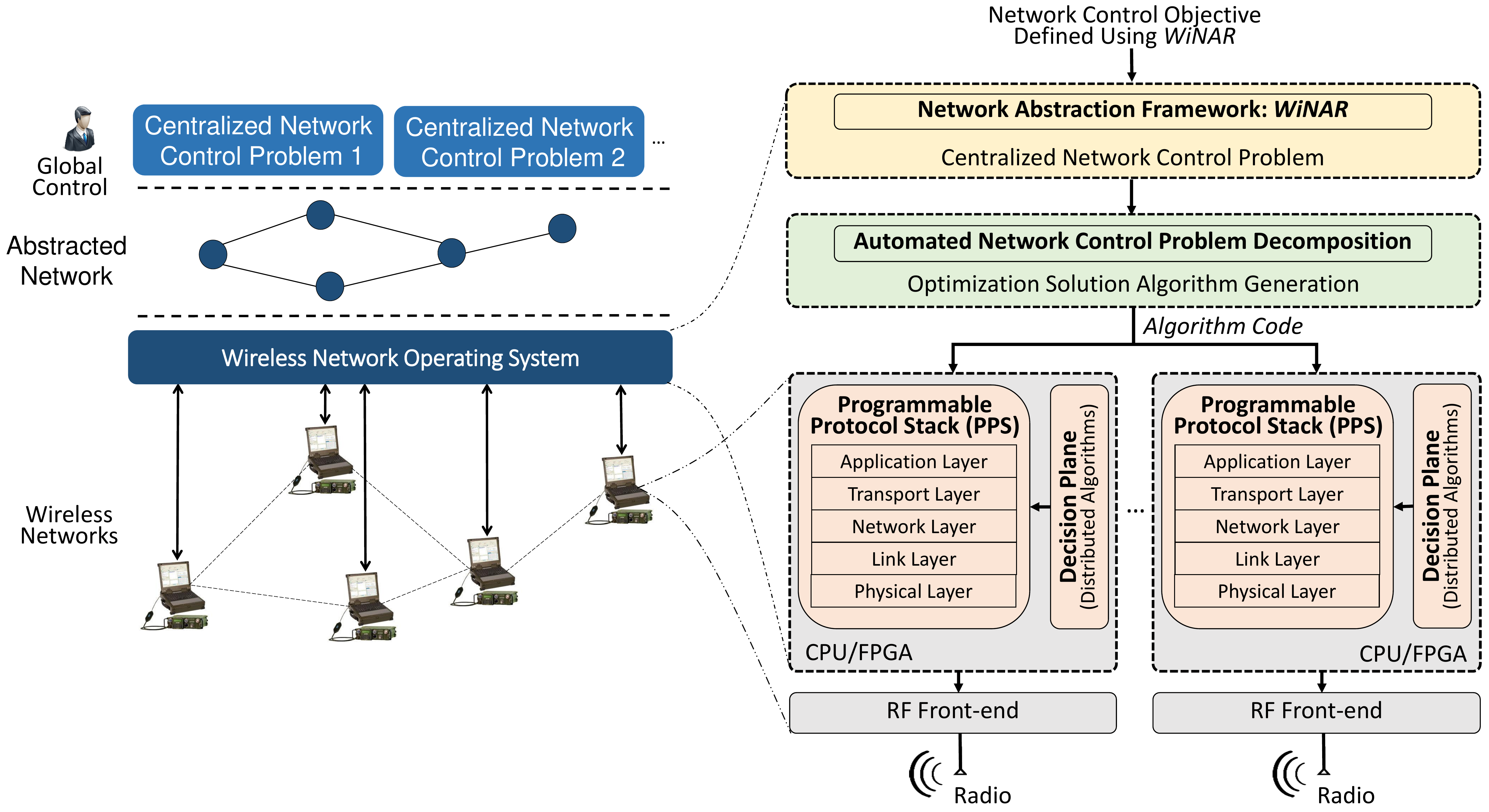} \vspace{-3mm}\caption{ \small Architecture of the wireless network operating system.}\vspace{-3mm}
\label{fig:arch}
\end{figure*}

\section{Introduction} \label{sec:intro}
Most existing wireless networks are inherently {hardware-based} and rely on closed and {\em inflexible architectures} that delay adoption of new wireless networking technologies. Moreover, it is very challenging to control large-scale networks of heterogeneous devices with diverse capabilities and hardware. Quite the opposite, software-defined radios provide a vast degree of flexibility. At the same time, software radios today lack appropriate abstractions to enable prototyping of complex networking applications able to leverage the cross-layer interactions that characterize wireless operations. To use an analogy from computer systems, trying to build a complex networked application on software radios is today as hard as trying to build a complex piece of enterprise software by writing bare-metal code in a low-level programming language.

There has been no lack of efforts trying to define new networking abstractions in recent years. The notion of software defined networking (SDN) has been introduced to simplify network control and to make it easier to introduce and deploy new applications and services as compared to classical hardware-dependent approaches \cite{OpenFlow, Akyildiz14, Mehran15}. The main ideas are (i) to separate the data plane from the control plane (an idea that in different form was already pervasive in the cellular industry); and more importantly (ii)
to ``control'' the network behavior through a centralized programmatic network abstraction. This simplifies the definition of new network control functionalities, which are now defined based on an {\em abstract} and {\em centralized} representation of the network.

So far, most SDN work has concentrated on ``softwarization'' of routing for commercial infrastructure-based wired networks, with some recent work addressing wireless networks \cite{OpenFlow, Bansal12, Gudipati13, YapPoster09, Erran12, Akyildiz14}. However, {\em applications of software-defined networking concepts to infrastructure-less wireless networks (i.e., tactical ad hoc networks, mesh, sensor networks, D2D, IoT) are substantially unexplored.\footnote{We will discuss a few exceptions in Section~ \ref{sec:related}: Related Work.}} This is not without a reason.
    Essentially, distributed control problems in wireless networks are complex and hard to separate into basic, isolated functionalities (i.e., layers in traditional networking architectures). Typical control problems in wireless networks involve making resource allocation decisions at multiple layers of the network protocol stack that are inherently and tightly coupled because of the shared wireless radio transmission medium; conversely, in software-defined commercial wired networks one can concentrate on routing at the network layer in isolation. Moreover, in most current instantiations of this idea, SDN is realized by (i) removing control decisions from the hardware, e.g., switches, (ii) by enabling hardware (e.g., switches, routers) to be remotely programmed through an open and standardized interface, e.g., OpenFlow \cite{OpenFlow}, and (iii) by relying on a centralized network controller to define the behavior and operation of the network forwarding
infrastructure. This unavoidably requires a high-speed backhaul infrastructure to connect the edge nodes with the centralized network controller, which is typically not available in wireless networks where network nodes need to make distributed, optimal, cross-layer control decisions at all layers to maximize the network performance while keeping the network scalable, reliable, and easy to deploy \cite{ZhaiTMC06, ToNMobiCom}. Clearly, these problems, which are specific to wireless, cannot be solved with existing SDN approaches.  

{\bf New Approach to Wireless SDN.} For these reasons, in this paper we propose and explore {\em a new approach to software-defined networking for  wireless networks}. At the core, we attempt to answer the following question: is it possible to {\em automatically generate distributed wireless network control programs} that are {\em defined} based on a centralized abstraction of the network that hides low-level implementation details; and in this way  \emph{bridge the gap between software defined networking and distributed network optimization/control?}
Can we, in this way, keep the benefits of distributed network control (where decisions are taken close to the network/channel/interference state without the need for collecting information at a centralized decision making point); and at the same time be able to define the network behavior based on a centralized abstraction? Can we, by answering these questions,  develop a {\em principled approach} to software-defined wireless networking based on cross-layer optimization theory? We attempt to provide a preliminary answer to these compelling questions by studying the core building principles of a \emph{Wireless Network Operating System (WNOS)}. Similar to a computer operating system, which provides the programmer with an abstraction of the underlying machine that hides the lower level hardware operations (e.g., its parallel nature in multi-core systems) and exposes only critical functionalities, WNOS provides the network designer with an abstraction hiding the lower-level details of the network operations. Maybe more importantly, WNOS hides the details of the {\em distributed implementation} of the network control operations, and provides the network designer with a centralized view abstracting the network functionalities at a high level. Based on this abstract representation, WNOS takes centralized network control programs written on a centralized, high-level view of the network and automatically generates distributed cross-layer control programs based on distributed optimization theory that are executed by each individual node on an abstract representation of the radio hardware.
This paper takes a decisive step in this direction and claims the following  contributions:
\begin{itemize}
\item \emph{WNOS Architecture Design}. We propose an architecture for WNOS by defining three key components: network abstraction, automated network control problem decomposition, and programmable protocol stack.
\item \emph{Network Abstraction}. We propose a new \underline{wi}reless \underline{n}etwork \underline{a}bstraction f\underline{r}amework \emph{WiNAR} - inspired by the language of network utility maximization (NUM), based on which network designers can characterize diverse desired network behaviors before actual deployment.
\item \emph{Automated Decomposition}. We propose the notion of \emph{disciplined instantiation}, based on which user-defined abstract centralized network control problems can be decomposed into a set of distributed subproblems in an automated fashion. Distributed control programs regulate the behavior of each involved node to obtain the desired centralized behavior in the presence of time-varying local conditions (including channel, traffic, etc.).
\item \emph{WNOS Prototyping and Testbed Evaluation}. We outline the design of a WNOS prototype that implements the proposed network abstraction and automated decomposition and solution algorithm generation approach, as well as a newly designed general purpose programmable protocol stack (PPS) that covers all protocol layers. Based on the PPS, a multi-hop wireless ad hoc network testbed is developed using software-defined radios to provide a proof of concept of the WNOS.
\end{itemize}

Unlike traditional SDN, which relies on centralized control (unsuitable for infrastructure-less wireless networks), we propose to {\em define control behaviors} on a centralized network abstraction, while {\em executing the behaviors} through automatically generated distributed control programs based on local network state information only. Hence, the user-defined centralized cross-layer network control objective can be achieved with no need to distribute network state at all layers of the protocol stack across the global network, which is obviously undesirable. We envision that the resulting WNOS will contribute to bridging the gap between centralized/distributed optimization techniques and software-defined networking - distributed control is not based on design-by-intuition principles, but on rigorous mathematical models based on nonlinear optimization theory.



The remainder of the paper is organized as follows. In Section~\ref{sec:arc}, we present the design architecture of WNOS, and then describe the network abstraction framework WiNAR in Section~\ref{sec:abst}. We discuss the automated network control problem decomposition approach in Section~\ref{sec:autodecomp}, and present the prototyping and experimental evaluation of WNOS in Sections~\ref{sec:testbed} and \ref{sec:evaluation}, respectively. We discuss limitations and future work in Section~\ref{sec:future} and review related work in Section~\ref{sec:related}.
Finally, we draw the main conclusions in Section~\ref{sec:conclusion}.


\vspace{-4mm}
\section{WNOS Architecture} \label{sec:arc}
The architecture of the proposed wireless network operating system (WNOS) is illustrated in Fig.~\ref{fig:arch}. At a high level, the WNOS comprises three key components: network abstraction, network control problem decomposition, and programmable protocol stack (PPS). 

\textbf{Network Abstraction}. This is the interface through which the network designer can define the network control problem to achieve certain application-specific objectives. 
Two core functionalities are provided by this component, that is, \emph{network behavior characterization} and \emph{centralized network control problem definition}. WNOS provides the designer with a rich set of network abstraction APIs through which the designer can characterize at a high-level the desired network behavior. Through the API, the designer can define various network control objectives, such as throughput maximization, energy efficiency maximization, delay minimization, or their combinations; can impose different constraints on the underlying physical network, such as the maximum transmission power of each node, available spectrum bandwidth, maximum end-to-end delay, among others. Importantly, to define a network control problem, the designer does not have to consider all implementation details of the networking protocol stack. That is, the designer can select different templates of network protocols, which are programmable with parameters that can be optimized in real time, such as deterministic scheduling vs stochastic scheduling, proactive routing vs reactive routing vs hybrid routing, delay-based vs packet-loss-based congestion control, among others.

It is worth pointing out that the network designer does not need to control protocol parameters manually. Instead, the parameters are optimized by WNOS through automatically generated distributed algorithms. These control objectives, network constraints, and selected protocol templates together serve as the input of the network control problem definition.
Then, given a network control problem defined at a high-level, a mathematical representation of the underlying centralized network utility maximization problem is constructed by parsing the network abstraction functions. Details of the network abstraction design will be discussed in Section~\ref{sec:abst}. 

\textbf{Network Control Problem Decomposition}. The resulting centralized network control problem, which characterizes the behavior of the wireless network, is then decomposed into a set of distributed sub-problems, each characterizing the local behavior, e.g., a single session or a single node. 
To this end, WNOS first determines a decomposition approach based on the mathematical structure of the network control problem, including whether the problem involves one or multiple sessions, what protocol layers are to be optimized, if the problem is convex or not, among others. Different decomposition approaches can lead to different structures of the resulting distributed control program with various convergence properties, communication overhead, and achievable network performance \cite{LOD,DPLJournal}.

Through vertical decomposition, a centralized network control problem can be decomposed into subproblems each involving a single or subset of protocol layers, while through horizontal decomposition each of the resulting subproblems involves local functionalities of a single session or node device. Different decomposition approaches can be jointly and iteratively applied if the centralized network control problem involves multiple concurrent sessions and cross-layer optimization of multiple protocol layers.
For each of the resulting subproblems, a numerical solution algorithm (e.g., interior-point method) is then selected to solve the problem. Different distributed solution algorithms interact with each other by updating and passing a common set of optimization variables. See Section~\ref{sec:decomp} for details of the decomposition approach. 

\textbf{Programmable Protocol Stack (PPS)}. For each of the resulting distributed network control problems, a numerical solution algorithm is selected to solve the optimization problem. This is executed in real time and the obtained optimization results are used to configure the control parameters of a PPS on each local network device to adapt to the dynamic networking environments.
The PPS provides abstractions and building blocks necessary to prototype complex cross-layer protocols
based on a high level, abstract representation of the software radio platform without
hiding, and instead while retaining control of, implementation details at all layers of the protocol stack and while maintaining platform independence \cite{openradio, WMAC, RcUBe}. The control interface between the PPS and the distributed solution algorithms is defined so that (i) the solution algorithm can retrieve network status information from the register plane of the PPS, such as noise and interference power level, queue status, available spectrum band, among others, and then use the retrieved information as input parameters of the distributed optimization problems; and (ii) based on the optimized solutions, the programmable protocol stack is able to configure in an on-line fashion the parameters of the adopted protocol stack via its decision engine in the decision plane, e.g., update the modulation scheme based on the optimized transmission power hence SINR, configure the TCP window size based on the optimized application-layer rate injected into the network.



\vspace{-2mm}
\section{Network Abstraction: WiNAR} \label{sec:abst}
The objective of the network abstraction component WiNAR is to provide network designers with an interface to characterize network behaviors at a high and centralized level. This goal is however not easy to accomplish because of the following main challenges:
\begin{itemize}
\item \emph{Pre-deployment network abstraction}. Unlike traditional network abstraction and resource virtualization \cite{Chengchao15, Abdeltouab12}, where the objective is to abstract or virtualize networks at one or two protocol layers at run time with fixed network topology and known global network information, in our case run-time information is not available in the design phase. For example, the available links that can be used by a session, the neighbors or interferers of a node, among others are not known a-priori. Therefore, the challenge is to abstract the wireless network before actual deployment by taking run-time uncertainties at all protocol layers into consideration, including time-varying wireless channels, interference coupling among nodes, network topology and traffic load variations, among others.
\item \emph{Multi-role network element}. A physical network entity may serve in different roles in the network. For example, a node can be the source or destination of a session, the transmitter, relay or receiver of a link, the neighbor of other nodes, a head of a cluster, a member of the whole network, among others. The network abstraction needs to allow designers to characterize network element behaviors with respect to heterogeneous roles while controlling the same physical network entity.
\end{itemize}

\noindent To address these challenges, elements in WNOS are represented following a three-fold abstraction.
At the core of the network abstraction there is a \emph{network representation layer}, which bridges the outer \emph{network control interface layer} and inner \emph{network modeling layer}. Through the network control interface layer, the designer defines the network control objective at a high level, and a mathematical representation of the defined centralized network control problem is then constructed based on the network modeling layer.



\textbf{Network Representation}.
The network abstraction represents different network entities as two categories of network elements, i.e.,  \emph{Primitive Element} and \emph{Virtual Element},  defined as follows.

\vspace{-2mm}
\begin{definition}[Primitive Element] \label{def:prmelmt} A primitive element is a network element that represents an individual determined network entity. Two criteria need to be satisfied for each primitive element~$\mathcal{A}$:
\begin{itemize}
\item $ |\{\mathrm{Network\; entities\; represented\; by\;} \mathcal{A}\}| = 1$ with $|\cdot|$ being the cardinality of a set, i.e., there exists a one-to-one mapping between any primitive element and a physical network entity.
\item For any time instants $t_1 \neq t_2$, $\mathcal{A}(t_1) = \mathcal{A}(t_2)$ always holds, i.e., the one-to-one mapping does not change with time.
\end{itemize}
\end{definition}
\noindent Examples of primitive elements include $Node$, $Link$, $Session$, $Link$ $Capacity$ and $Session\; Rate$, among others.\footnotemark\footnotetext{Here, $Link\; Capacity$ and $Session\; Rate$ refer to the network parameters rather than any specific values of the parameters that can be time varying.}

\vspace{-2mm}
\begin{definition}[Virtual Element] \label{def:vtlelmt} A virtual element represents an undetermined set of network entities, i.e., cannot be mapped to a deterministic set of primitive elements other than at runtime
A virtual element $\mathcal{V}$ satisfies
\begin{itemize}
\item $ |\{\mathrm{Network\; entities\; represented\; by\;} \mathcal{V}\}| \geq 1$, i.e., each virtual element is mapped to physical network entities in a one-to-many manner.
\item $\mathcal{V} = \mathcal{V}(t)$, i.e., the set of network entities represented by each virtual element is a function of the network run time~$t$.
\end{itemize}
\end{definition}
\vspace{-2mm}
\noindent Examples of virtual element include $Neighbors\; of\; Node$ (the set of neighbors of a node), $Links\;of\;Session$ (the set of links used by a session), $Sessions\; of\; Link$ (the set of sessions sharing the same link), among others. The members of a virtual element are primitive elements, e.g., each member of virtual element $Links\;of\;Session$ is a primitive element $Link$.

\begin{figure}[t]
\centering
\includegraphics[width=0.4\textwidth]{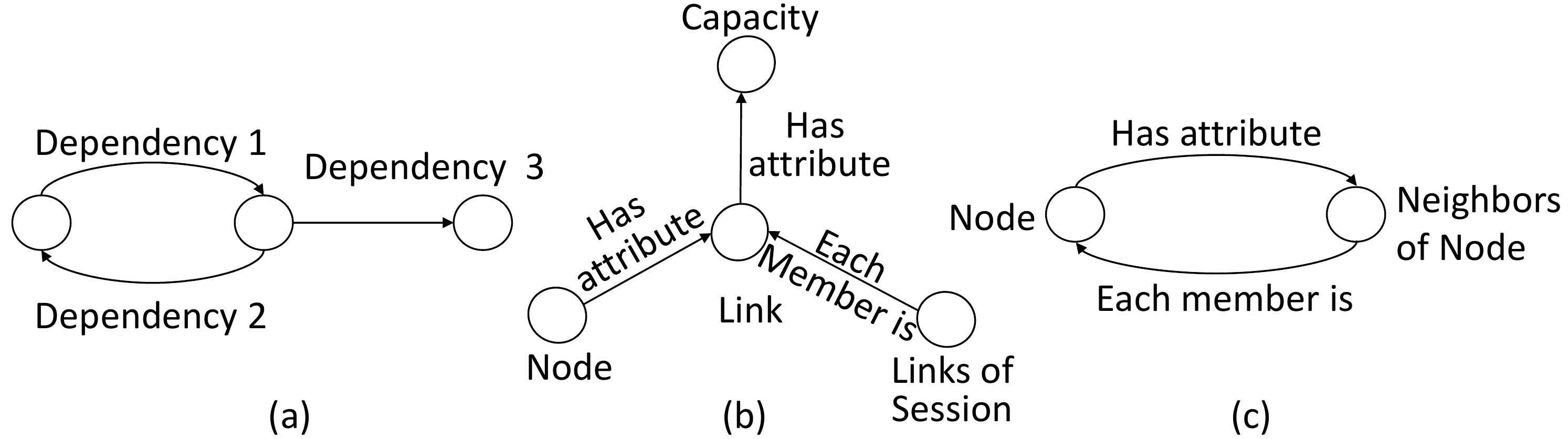} \vspace{-3mm} \caption{\small Representation of the dependency relationship among network elements based on directed multigraph. (a): Illustration of directed multigraph; (b) and (c): Graph examples.\vspace{-5mm}}
\label{fig:graph}
\end{figure}



Then, a wireless network can be characterized using a set of primitive and virtual network elements as well as the cross-dependency among the elements, which is formalized in \emph{Definition}~\ref{def:net}.
\vspace{-2mm}
\begin{definition}[Network] \label{def:net} With primitive elements $\mathcal{A}_m, \mathcal{A}_{m'}$ and virtual elements $\mathcal{V}_{n}, \mathcal{V}_{n'}$,
a network $Net$ can be represented as
\begin{align} \label{eq:net}
Net = \{ & \mathcal{A}_m, \mathcal{V}_n, I(\mathcal{A}_m, \mathcal{A}_{m'}), I(\mathcal{V}_n, \mathcal{V}_{n'}), I(\mathcal{A}_m, \mathcal{V}_{n})\;  \nonumber\\
&m, m'\in \mathcal{M}_A, m\neq m',  n, n' \in\mathcal{N}_V, n\neq n'\}
\end{align}
where $\mathcal{M}_A$ and $\mathcal{N}_V$ are the sets of primitive and virtual network elements, respectively, and $I(\mathcal{A}_m, \mathcal{A}_{m'}), I(\mathcal{V}_n, \mathcal{V}_{n'}), I(\mathcal{A}_m, \mathcal{V}_{n})$ represent the inter-dependencies between primitive elements $\mathcal{A}_m$ and $\mathcal{A}_{m'}$, between virtual elements $\mathcal{V}_n$ and $\mathcal{V}_{n'}$, between primitive element $\mathcal{A}_m$ and virtual element $\mathcal{V}_n$, respectively.
\end{definition}

\vspace{-1mm}
In \emph{Definition}~\ref{def:net}, the inter-dependency $I(\cdot, \cdot)$ among different network elements can be characterized as a directed multigraph \cite{Balakrishnan97}.
As illustrated in Fig.~\ref{fig:graph}(a), each vertex of the graph represents a network element, and the relationship between two coupled vertices are characterized using one or multiple directed edges connecting the two vertices. All directed edges together characterize the cross-dependency relationship among the network elements. Figures \ref{fig:graph}(b) and (c) are two examples of the multigraph-based network element representation. In Fig.~\ref{fig:graph}(b), primitive element ${Link}$ is the holder of another primitive element ${Capacity}$ (i.e., ${Link}$ has attribute ${Capacity}$). Similarly, ${Link}$ is an attribute of primitive element ${Node}$ and is a member of virtual element ${Links\; of\; Session}$. In Fig.~\ref{fig:graph}(c), the mutual relationship between primitive element ${Node}$ and virtual element ${Neighbors\; of\; Node}$ are characterized using two directed edges (hence a multigraph \cite{Balakrishnan97}): ${Node}$ has an attribute ${Neighbors\; of\; Node}$, each member of which is a $Node$.



\begin{itemize}
\item \emph{Has Attribute} characterizes parent-child relationships between network elements, e.g., parent element $Link$ has child elements $Link\; Capacity$ (lnkcap) and $Link\; Power$ (lnkpwr) as its attributes.
\item \emph{Each Member is} characterizes set-individual relationships between virtual and primitive elements, e.g., each member of $Links \; of\; Session$ (lnkses) is a $Link$ (lnk).
\item  \emph{Is Function of} defines the mathematical model of an element based on other elements, e.g., element $Link\; SINR$ (lnksinr) is a function of $Link\;Power$ (lnkpwr).
\end{itemize}

\vspace{-1mm}
\textbf{Network Control Interfaces}. Based on the network element representation, network control interfaces can then be designed. Based on these, network designers are allowed to characterize network behaviors. Four categories of operations have been defined:
\begin{itemize}
\item \emph{Read}: Extract network information from a single or a group of network elements, e.g., extract the set of links used by a session from the attributes of $Node$.
\item \emph{Set}: Configure parameters for a single or a group of network elements, e.g., set $Maximum\; Power$ (i.e., maxpwr), which is also an attribute of element $Node$.
\item \emph{Compose}: Construct a new expression by mathematically manipulating network parameters obtained through $Read$ operations. For example, add together the power of all links originated from the same node, i.e., sum $Link\;Power$ (lnkpwr) over $Links\;of\;Node$ (lnknd).
\item \emph{Compare}: Define network constraints by comparing two expressions obtained using \emph{Compose} operations.
\end{itemize}




\textbf{Centralized Network Control Problem}. Finally, centralized network control problems can be defined based on the network control interfaces.
A network control problem comprises of four components: network setting, control variables, network utility and network constraints.
\begin{itemize}
\item \emph{Network Setting} can be configured by setting network parameters using \emph{Set} operations and extracted from network elements using \emph{Read} operations. Configurable network parameters include network architecture (single- or multi-hop, flat or clustered topology), spectrum access preferences (scheduled or statistical access), routing preferences (single- or multi-path routing), among others.
\item \emph{Control Variables} can be defined by setting (i.e., \emph{Set} operation) network parameters as optimization variables, including transmission power, frequency bandwidth, transmission time, source rate, channel access probability, among others.
\item \emph{Network Utility} can be defined by binding (i.e., \emph{Compose} operation) one or multiple expressions with mathematical operations like $+$, $-$, $\times$, $\div$ and mathematical functions like $\log$, $\sqrt{(\cdot)}$ and their combinations.
\item \emph{Network Constraints} can be defined by comparing two expressions using \emph{Compare} operations. 
\end{itemize}

\vspace{-1mm}
\noindent Examples of network control problem definition based on the developed abstraction APIs will be given in Section~\ref{sec:evaluation}.

Given the high-level characterization of network behaviors, the underlying mathematical models of the problem can then be constructed by extracting the mathematical models of each network element using the \emph{Read} operation. 
The resulting network utility maximization problem is a centralized cross-layer network optimization problem. Our goal is to generate, \emph{in an automated fashion}, distributed control programs that can be executed at individual network devices, which is accomplished by another main component of WNOS, i.e., \emph{Network Control Problem Decomposition} as described in Section~\ref{sec:autodecomp}.


\vspace{-2mm}
\section{Automated Network Control Problem Decomposition} \label{sec:autodecomp}
So far, there is no existing unified decomposition theory that can be used to decompose arbitrary network control problems. Depending on whether we need to decompose coupled network constraints, or coupled radio resource variables; and depending on the decomposition order, a cross-layer network control problem can be theoretically decomposed based on dual decomposition, primal decomposition, indirect decomposition and their combinations. Please refer to \cite{LOD, Daniel06, Johansson06,DPLJournal} and references therein for a tutorial and survey of existing decomposition theories and their applications. In this paper, as one of our major contributions, we propose an automated network control problem decomposition approach based on decomposition of nonlinear optimization problems.

The core objective of the decomposition is two-fold:
\vspace{-1mm}
\begin{itemize}
\item \emph{Cross-layer Decomposition}: Decouple the coupling among multiple protocol layers, resulting in subproblems each involving functionalities handled at a single protocol layer;
\item \emph{Distributed Control Decomposition}: Decouple the coupling in radio resource allocation among different network devices, resulting in subproblems that can be solved at each device in a distributed fashion.
\end{itemize}
\vspace{-1mm}
Next, we first provide a brief review of  cross-layer distributed decomposition theory based on which our automated decomposition approach is designed.


\vspace{-2mm}
\subsection{Decomposition Approaches}\label{sec:theory}
\textbf{Cross-layer Decomposition}. In this paper we consider duality theory for cross-layer decomposition (while the automated decomposition approach in Section~\ref{sec:decomp} is not limited to any specific decomposition theory). Consider a network control problem expressed as
\vspace{-2mm}
\begin{eqnarray}\label{eq:original}
\begin{array}{cl}
\underset{\mathbold{x}}{\mathrm{maximize}} & \sum\limits_{i\in\mathcal{I}} f_i(x_i), \\
\mathrm{subject\;to:} & \sum\limits_{i\in\mathcal{J}_i} g_i(x_i) \leq c_j, \; \forall j\in \mathcal{J} \vspace{-2mm}
\end{array}
\end{eqnarray}
with $\mathbold{x} = (x_i)_{i\in\mathcal{I}}$ being the control vector. The dual function can be constructed by incorporating the constraints into utility in (\ref{eq:original}) by introducing Lagrangian variables $\mathbold{\lambda} = (\lambda_j)_{j\in\mathcal{J}}$,
\vspace{-1mm}
\begin{align}
\mathrm{maximize}\; L(\mathbold{x}, \mathbold{\lambda}) =  \sum_{i\in\mathcal{I}} f_i(x_i) - \sum\limits_{j\in \mathcal{J}} \lambda_j\left (c_j - \sum\limits_{i\in\mathcal{J}_i} g_i(x_i)\right)\label{eq:dual}
\end{align}
where $L(\mathbold{x}, \mathbold{\lambda})$ is called the Lagrangian of problem (\ref{eq:original}) \cite{Boyd04}. Then, the original problem (\ref{eq:original}) can be solved in the dual domain by minimizing (\ref{eq:dual}), i.e.,  minimizing the maximum of the Lagrangian. This can be accomplished by decomposing (\ref{eq:dual}) into subproblems
\vspace{-2mm}
\begin{align}
& f_{\mathrm{sub\_1}} = \underset{\mathbold{x}}{\mathrm{maximize}}\; \sum_{i\in\mathcal{I}} f_i(x_i) + \sum\limits_{j\in \mathcal{J}} \lambda_j\left ( \sum\limits_{i\in\mathcal{J}_i} g_i(x_i)\right),\label{eq:sub1}\\
& f_{\mathrm{sub\_2}} =  \underset{\mathbold{\lambda}}{\mathrm{minimize}}\;  f_{\mathrm{sub\_1}} - \sum\limits_{j\in \mathcal{J}} \lambda_j c_j, \label{eq:sub2}\vspace{-3mm}
\end{align}
and then iteratively maximizing $f_{\mathrm{sub\_1}}$ over control variables $\mathbold{x}$ with given $\mathbold{\lambda}$ and updating $\mathbold{\lambda}$ with the minimizer of $f_{\mathrm{sub\_2}}$.


\textbf{Distributed Decomposition}.
The outcome of cross-layer decomposition is a set of network control subproblems each corresponding to a single protocol layer, e.g., capacity maximization at the physical layer, delay minimization through routing at the network layer, among others.
The objective of distributed decomposition is to further decompose each of the resulting single-layer subproblems into a set of local network control problems that can be solved distributively at each single network entity based on local network information.

In the existing literature, this goal has been accomplished by designing distributed network control algorithms manually for specific network scenarios and control objectives \cite{JOPC05, Chen06, LDing10TVT, Yuan13, ToNMobiCom}, which however requires deep expertise in distributed optimization. Next, we present a theoretical framework based on which distributed control programs can be designed for \emph{arbitrary} user-defined network control problems.

The core design principle is to decompose a coupled multi-agent network control problem into a set of single-agent subproblems, where each agent optimizes a penalized version of their own utility. Consider a multi-agent network control problem with the objective of $\mathrm{maximizing}\; U(\mathbf{x}) \triangleq \sum\limits_{i\in\mathcal{I}} U_i(\mathbf{x}_i, \mathbf{x}_{-i})$,
where $U_i$ is the utility function of agent $i\in\mathcal{I}$, $\mathbf{x} = (\mathbf{x}_i, \mathbf{x}_{-i})$ with $\mathbf{x}_i$ and $\mathbf{x}_{-i}$ representing the strategy of agent $i$ and the strategy all other agents in $\mathcal{I}/i$. Then, the key of distributed decomposition is to construct a penalized individual utility $\widetilde{U}_i(\mathbf{x}_i, \mathbf{x}_{-i})$ for each agent $i\in\mathcal{I}$, expressed as $\widetilde{U}_i(\mathbf{x}_i, \mathbf{x}_{-i}) = \Theta_i(U(\mathbf{x})) + {\Gamma}_i(\mathbf{x})$,
where $\Theta_i(U(\mathbf{x}))$ is the individual item of $U(\mathbf{x})$ associated to agent $i\in\mathcal{I}$, ${\Gamma}_i(\mathbf{x})$ is the  penalization item for agent $i$. Below are three special cases of $\widetilde{U}_i(\mathbf{x}_i, \mathbf{x}_{-i})$ while both individual and penalization items can be customized by network designers to achieve a trade-off between communication overhead and social optimality of the resulting distributed control programs.

\begin{itemize}
\item Case 1: $\Theta_i(U(\mathbf{x})) = f_i(\mathbf{x}_i, \mathbf{x}_{-i})$, ${\Gamma}_i(\mathbf{x}_i, \mathbf{x}_{-i}) = 0$, i.e., \emph{best response without penalization}. In this case, the agents optimize their own original utility $U(\mathbf{x}_i, \mathbf{x}_{-i})$ by computing the best response to the strategies of all other competing agents (i.e., $\mathbf{x}_{-i}$) with zero signaling exchanges.
\item Case 2: $\Theta_i(U(\mathbf{x})) = \nabla_{\mathbf{x}_i} U_i(\mathbf{x}^0) (\mathbf{x}_i - \mathbf{x}_i^0)$, ${\Gamma}_i(\mathbf{x}) = \sum\limits_{j\in\mathcal{I}/i} \nabla_{\mathbf{x}_i} U_j(\mathbf{x}^0)$ $(\mathbf{x}_i - \mathbf{x}_i^0)$, with $\mathbf{x}_i^0$ and $\mathbf{x}^0$ being the current strategy of agent $i$ and of all agents. This will result in distributed gradient algorithm \cite{Dimitri97}, where partial cooperation is allowed among the agents by exchanging appropriate signaling messages.
\item Case 3: $\Theta_i(U(\mathbf{x})) = U_i(\mathbf{x}_i, \mathbf{x}_{-i}^0)$,  ${\Gamma}_i(\mathbf{x}_i, \mathbf{x}_{-i}) = \sum\limits_{j\in\mathcal{I}/i} \nabla_{\mathbf{x}_i} f_j(\mathbf{x}^0)$, which leads to decomposition by partial linearization (DPL), a newly established decomposition result \cite{DPLJournal}.
\end{itemize}


\begin{figure}[b]
\centering
\vspace{-5mm}
\includegraphics[width=0.4\textwidth]{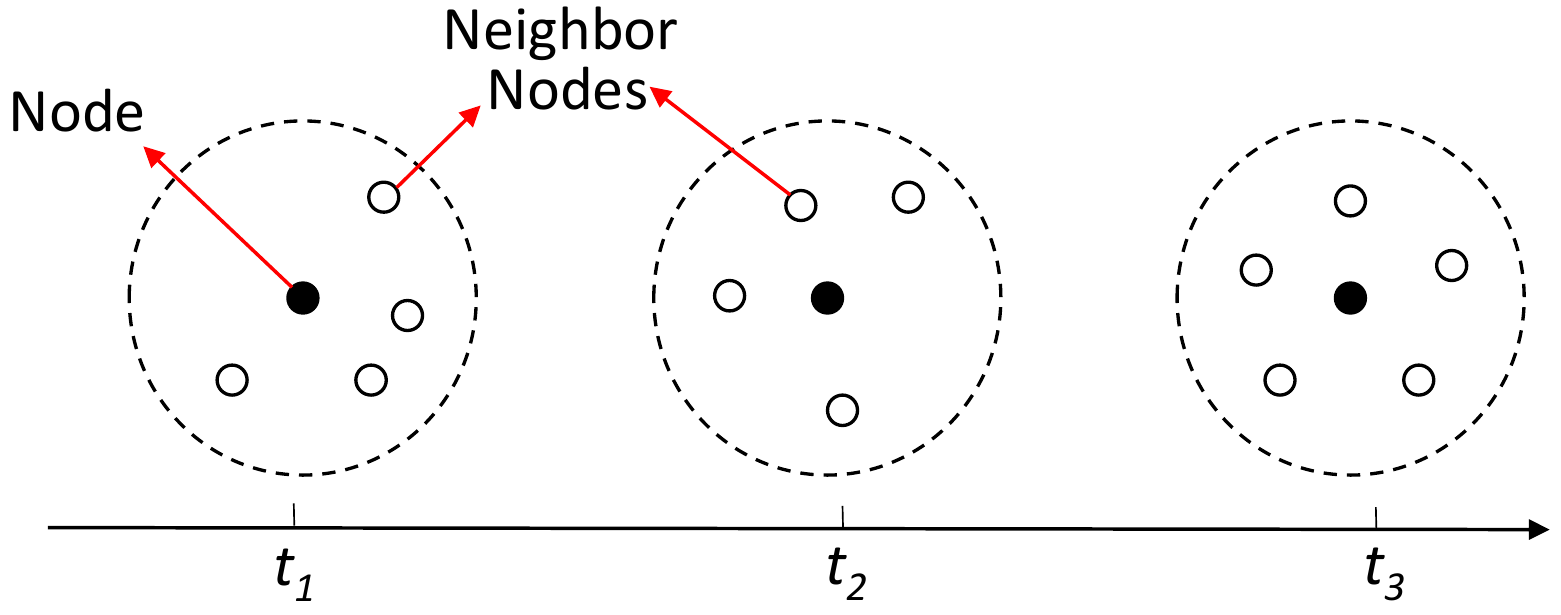} \vspace{-5mm}\caption{\small Illustration of time-varying set of neighbors nodes.\vspace{-5mm}}
\label{fig:ndnbr}
\end{figure}

\vspace{-2mm}
\subsection{Automated Decomposition} \label{sec:decomp}


A key step in cross-layer decomposition, as discussed in Section~\ref{sec:theory}, is to form a dual function for the original user-defined network control problem by absorbing constraints into the utility. Here, an underlying assumption is that the original problem (\ref{eq:original}) must have a determined set of constraints, i.e., sets $\mathcal{I}$, $\mathcal{J}$ and $\mathcal{J}_i, \forall i\in\mathcal{I}$ in (\ref{eq:original}) must be known. This poses significant challenges to automated network control problem decomposition at design phase, because the sets associated to the network elements are not determined \emph{other than at run time}, i.e., they are virtual elements as defined in Section~\ref{sec:abst}.

Take virtual element \emph{nbrnd} as an example, i.e., the set of $Neighbors\;$ $of\; Node$. As illustrated in Fig.~\ref{fig:ndnbr}, the neighbors of a node may change from time to time because of movement of nodes, joining of new nodes or leaving of dead nodes. Similarly, the set of links along an end-to-end path, the set of sessions sharing the same link and the set of all active links in the network, among others, are also time varying with no predetermined sets. That is to say, a network control problem defined at a high and abstract level may result in many instances of problems with different sets in the constraints and hence different dual variables $\lambda_j$ in the resulting dual function (\ref{eq:dual}). Therefore, \emph{a centralized user-defined network control problem cannot be decomposed by decomposing an arbitrary specific instance of the problem}.

As a core contribution of this work, next we present a new methodology to enable network control problem decomposition in an automated fashion at design phase with no need to know run time network information. At the core, we ask the following question: \emph{For a user-defined centralized abstract network control problem, are there any special set of instances of the problem such that decomposing any problems in the special set decomposes all possible instances? If yes, what is the right approach to obtain such problem instances?} We answer these questions by proposing the notion of \emph{disciplined instantiation (DI)}.


\begin{figure}[b]
\centering
\includegraphics[width=0.45  \textwidth]{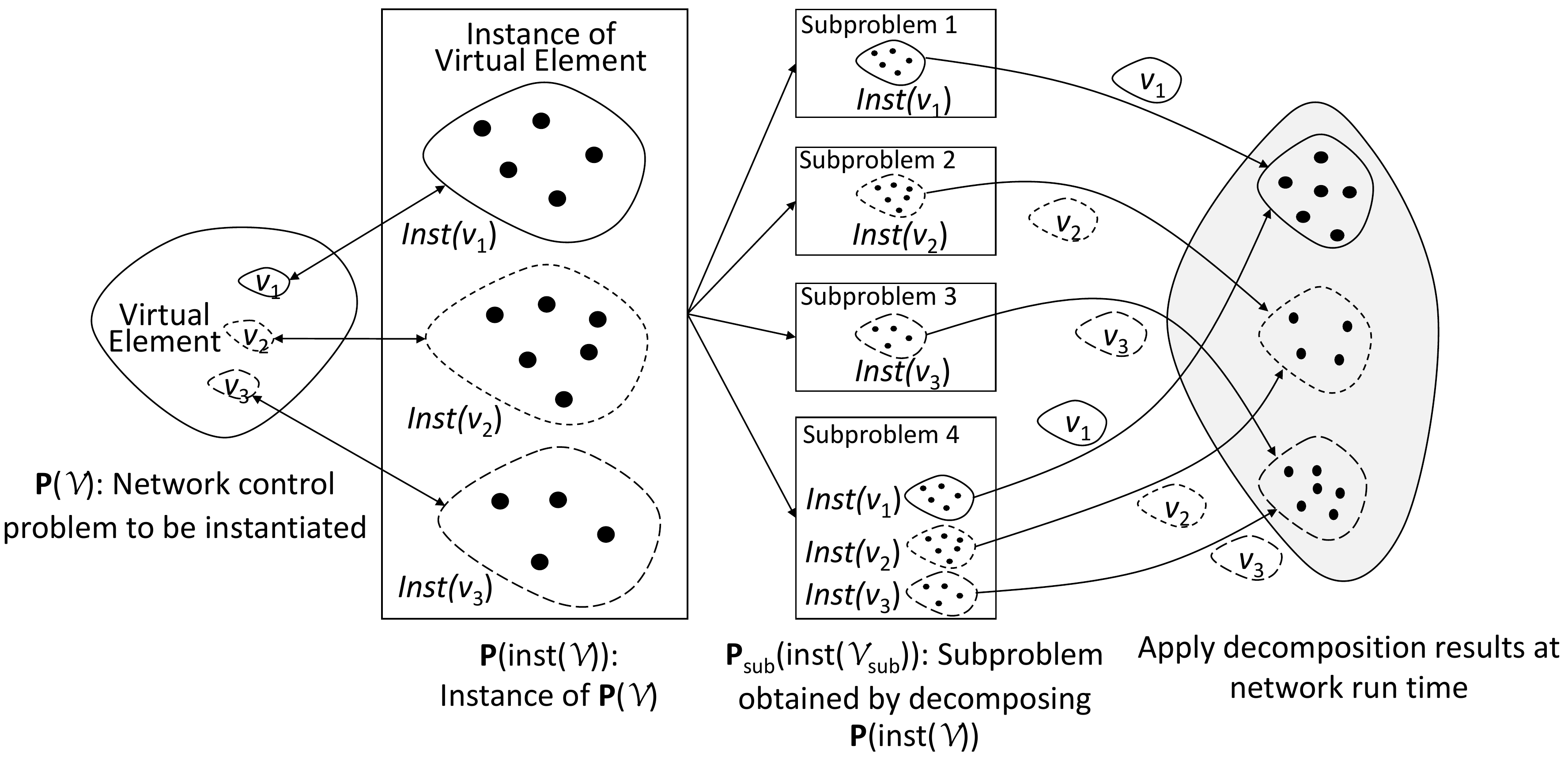} \vspace{-4mm}\caption{\small Basic principle of network control problem decomposition based on disciplined instantiation  (DI).\vspace{-5mm}}
\label{fig:revinst}
\end{figure}

\textbf{Disciplined Instantiation}. In a nutshell, the DI technique generates at design time, following certain rules (as discussed below),  a specific instance of the user-defined abstract network control problem, such that the abstract problem can be decomposed by decomposing the specific instance and the obtained decomposition results can be applied to those control problems at network run time.

In Fig.~\ref{fig:revinst} we illustrate the basic principle of the DI-based decomposition approach by considering a network control problem that involves three virtual elements $v_1$, $v_2$ and $v_3$, which, e.g., can be $Neighbors\;$ $of\; Node$ for nodes 1, 2 and 3, respectively.
Let $inst(v_i)$ represent the instance of virtual element $v_i$, denote $\mathcal{V} = \{v_1, v_2, v_3\}$ as the set of all the three virtual elements and further denote the set of instances for all $v_i \in \mathcal{V}$ as $inst(\mathcal{V})$.
Then, the objective of DI is to create a unique instance for each virtual element $v_i\in \mathcal{V}$ such that there exists a one-to-one mapping between $\mathcal{V}$ and $inst(\mathcal{V})$.

Denote $\mathbf{P}(\mathcal{V})$ as the network control problem to be instantiated, and let $\mathbf{P}(inst(\mathcal{V}))$ represent the specific instantiated problem obtained by instantiating  $\mathbf{P}(\mathcal{V})$. Then, $\mathbf{P}(inst(\mathcal{V}))$ can be decomposed into a set of subproblems $\mathbf{P}_{\mathrm{sub}}(inst(\mathcal{V_{\mathrm{sub}}}))$ each involving only a subset $\mathcal{V}_{\mathrm{sub}}$ of the virtual elements with  $\mathcal{V}_{\mathrm{sub}} \subset \mathcal{V}$. For example, in Fig.~\ref{fig:revinst}, $\mathbf{P}(inst(\mathcal{V}))$ has been decomposed into four subproblems, with the first subproblem involving only virtual element $v_1$, the second involving only $v_2$, the third involving only $v_3$ while the fourth involves all three virtual elements. Because of the one-to-one mapping between each virtual network element $v_i$ and its instance $inst(v_i)$, the decomposition results obtained by decomposing  $\mathbf{P}(inst(\mathcal{V}))$ are also applicable to the original problem $\mathbf{P}(\mathcal{V})$ represented in virtual elements and hence its various specific instances at network run time.


In the above procedure, the key is to guarantee one-to-one mapping between each virtual element $v_i$ and its instance $inst(v_i)$. This cannot be achieved by generating arbitrarily disjoint instances for different virtual elements $v_i$ because active network elements assume multiple roles as described in Section~\ref{sec:abst}. For example, a physical link needs to be involved in the instances of virtual element \emph{ ``Links of Session"} for all the sessions sharing the link. In the following, we first describe the two rules following which instances are generated in WNOS, i.e., \emph{equal cardinality} and \emph{ordered uniqueness}, and then discuss why the two rules are needed for DI. Before this, we first identify two categories of virtual elements, i.e., \emph{global} and \emph{local} virtual elements. Please refer to Section~\ref{sec:abst} for the definition of virtual element.
\begin{itemize}
\item A \emph{global virtual element} is a virtual element whose set of physical network entities have the same entity type (e.g., node, or link) and spans over the entire network, e.g., element \emph{netnd} represents $Nodes\;in\;Network$, the set of all users $\mathcal{I}$ in (\ref{eq:original})-(\ref{eq:sub1}).  
\item Differently, a \emph{local virtual element} comprises a subset of physical network entities of the network, and hence is a subset of the corresponding \emph{global virtual element}. For example, local virtual element nbrnd (i.e., $Neighbors\; of\; Node$) is a subset of global virtual element $Nodes\; in\; Network$; as another example, in (\ref{eq:original})-(\ref{eq:sub1}), since $\mathcal{J}_i$ is a subset of $\mathcal{J}$, i.e., $\mathcal{J}_i \subset \mathcal{J}$,  $\mathcal{J}_i$ is a local virtual element while $\mathcal{J}$ is a global virtual element.
\end{itemize}

\textbf{Rule 1: Equal Cardinality}. This rule requires that all the instances for the same type of local virtual elements, e.g., $Neighbors\; of$\; $Node$, must have the same cardinality, i.e., the same number of members. Instances that satisfy this requirement are called \emph{peer} instances.

In WNOS, this is achieved by \emph{peer random sampling}, a technique that can be used to generate \emph{peer} instances. Specifically, given a user-defined network control problem, the global virtual element denoted as $v^{\mathrm{glb}}$ is first instantiated using a set of pre-determined number $N^{\mathrm{glb}}$ of elements, i.e., $|inst(v^{\mathrm{glb}})| = N^{\mathrm{glb}}$ with $inst(v^{\mathrm{glb}})$ being the instance of the global virtual element $v^{\mathrm{glb}}$ and $|inst(v^{\mathrm{glb}})|$ being the cardinality of $inst(v^{\mathrm{glb}})$. The resulting instance $inst(v^{\mathrm{glb}})$ will be used to serve as the mother set to generate instances for those local virtual elements $v^{\mathrm{lcl}}$.

Then, each local virtual element  $v^{\mathrm{lcl}}$
can be instantiated by randomly selecting a subset of members from the mother set $inst(v^{\mathrm{glb}})$, i.e., the instance of the global virtual element $v^{\mathrm{glb}}$. Denote the resulting subset instance as $inst(v^{\mathrm{lcl}})$, then we have $|inst(v^{\mathrm{lcl}})| =  N^{\mathrm{lcl}}$ and $inst(v^{\mathrm{lcl}}) \subset inst(v^{\mathrm{glb}})$, where $N^{\mathrm{lcl}}$ is the cardinality of instances for local virutal elements.





\begin{figure}[b]
\centering
\vspace{-4mm}
\includegraphics[width=0.35\textwidth]{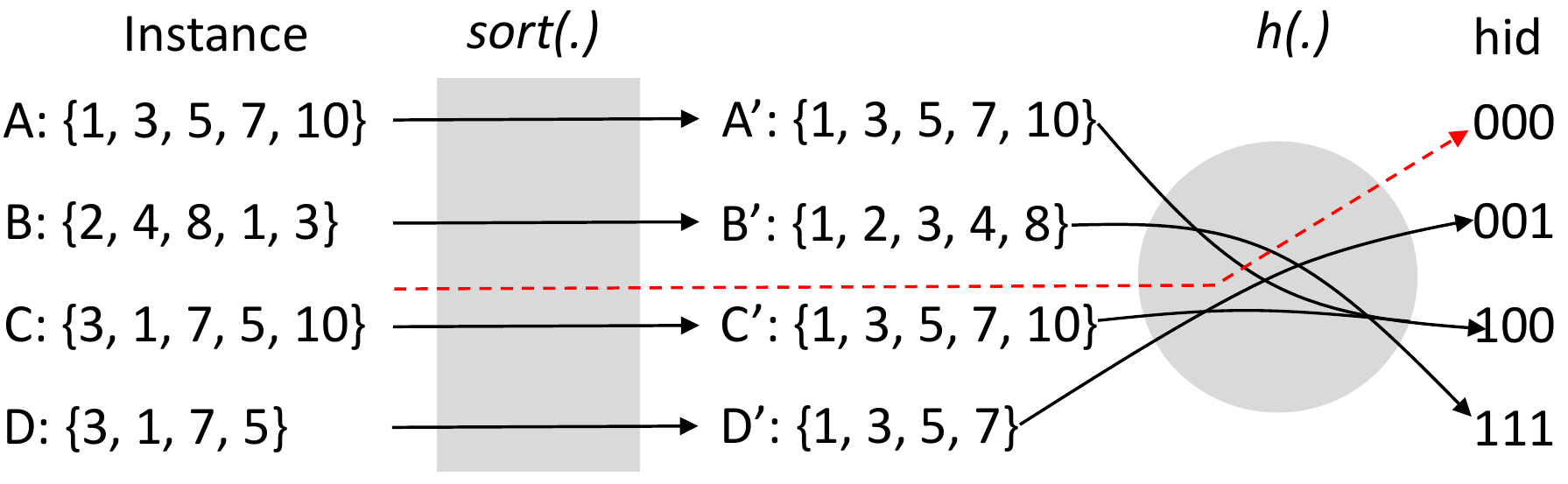} \caption{\small Illustration of hash mapping.}
\label{fig:hash}
\end{figure}
\textbf{Rule 2: Ordered Uniqueness}. With this rule, a unique instance will be generated for each local virtual element $v^{\mathrm{lcl}}$. This means that no two subsets generated by peer random sampling will be \emph{identical}.

In WNOS, this is accomplished by \emph{hash checking}, as in (\ref{eq:hash}):
\begin{equation}
inst(v^{\mathrm{lcl}}) \overset{sort(\cdot)}{\longrightarrow}  inst'(v^{\mathrm{lcl}}) \overset{h(\cdot)}{\longrightarrow} hid^{\mathrm{lcl}}, \label{eq:hash}
\end{equation}
where the members of $inst(v^{\mathrm{lcl}})$, i.e., the instance for local virtual element $v^{\mathrm{lcl}}$, are first sorted, and then
a unique id $hid^{\mathrm{lcl}}$ is calculated for the sorted instance $inst'(v^{\mathrm{lcl}})$ using a hash function $h(\cdot)$.
A hash function is a function that can be used to map an arbitrary-size data (instances in our case) to a fixed-size id \cite{hash77}. In WNOS, hash function is used to enable fast uniqueness checking by generating an id for each of the generated instances.


\textbf{Rationale for The Rules.}
The above two rules together guarantee that there exist a one-to-one mapping between the local virtual elements and their instances. As discussed above, this is the key to guarantee that the decomposition results obtained based on DI are also applicable at network run time. To show how the one-to-one mapping can be achieved following the two rules, we take  Fig.~\ref{fig:hash} as an example where A, B, C and D represent four specific instances of local virtual element $v^{\mathrm{lcl}}$, with each member in the set representing a primitive element (see Section~\ref{sec:abst} for the definition), e.g., an individual node.
Denote $\mathrm{A'}$, $\mathrm{B'}$, $\mathrm{C'}$ and $\mathrm{D'}$ as the sets resulting from sorting the members of $\mathrm{A}$, $\mathrm{B}$, $\mathrm{C}$ and $\mathrm{D}$, respectively.

It can be seen that set A is mapped to a three-digit id $\mathrm{100}$ while $\mathrm{B}$ is mapped to $\mathrm{111}$. Instance $\mathrm{C}$ is also mapped to $\mathrm{100}$ since its sorted instance $\mathrm{C}'$ is identical to $\mathrm{A'}$. Note that in Fig.~\ref{fig:revinst}, in each instantiated sub-problem $\mathbf{P}_{\mathrm{sub}}(inst(\mathcal{V}_{\mathrm{sub}}))$ the members of each instance may be re-ordered by the mathematical manipulations decomposing the instantiated network control problem $\mathbf{P}(inst(\mathcal{V}))$, e.g., forming and decomposing the dual function in (\ref{eq:dual}), (\ref{eq:sub1}) and (\ref{eq:sub2}). In (\ref{eq:hash}) function $sort(\cdot)$ guarantees that the same instances are always mapped to the same id regardless of the order of its members; otherwise, instance $\mathrm{C}$ will be mapped to a different id $\mathrm{000}$ as the red dashed arrow indicates in Fig.~\ref{fig:hash}.

\begin{figure}[t]
\centering
\includegraphics[width=0.2\textwidth]{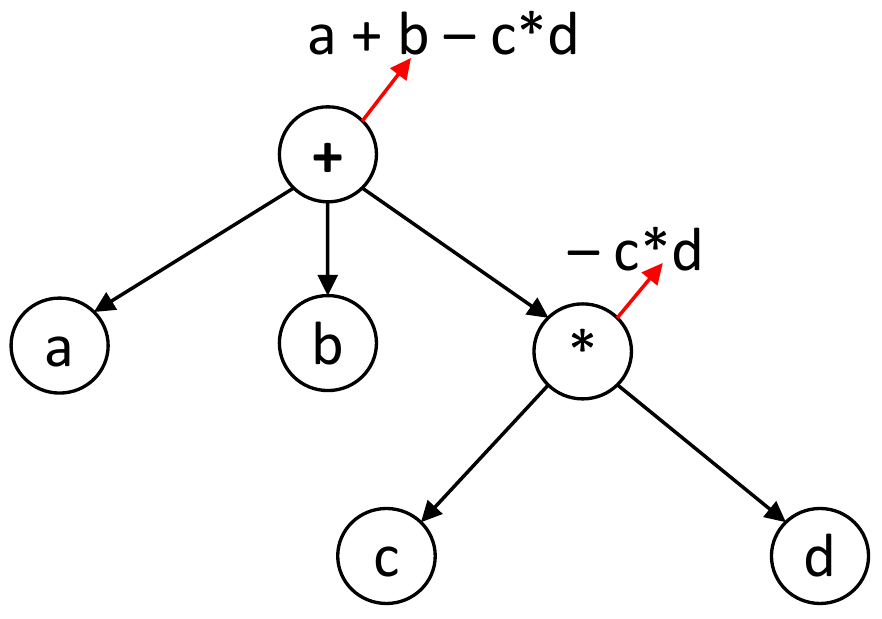} \vspace{-5mm}\caption{\small Tree representation of mathematical expressions.\vspace{-5mm}}
\label{fig:tree}
\end{figure}

Moreover, in Fig.~\ref{fig:hash} instance $\mathrm{D}$ is mapped to an id different from that of $\mathrm{A}$ and $\mathrm{C}$. This implies that an instance $\mathrm{A}$ and its subset instance $\mathrm{D}$ cannot be used at the same time for disciplined instantiation (DI); otherwise, it will be hard to separate them if they appear in the same instantiated network control sub-problems $\mathbf{P}_{\mathrm{sub}}(inst(\mathcal{V}_{\mathrm{sub}}))$. In DI, this is prevented by keeping all local instances \emph{peer}, i.e., it holds true for all local virtual elements that no instance is a proper subset of any other instances.
If hash checking finds that a new instance for local virtual element $v^{\mathrm{lcl}}$ is \emph{identical} to any existing instances, i.e., they have the same id, another instance will be created $v^{\mathrm{lcl}}$ by peer random sampling. 

Following the above two rules, a unique specific instance can be obtained for each of the virtual local elements, while there exists a one-to-one mapping relationship between the local virtual elements and their instances. Thus, decomposing the original network control problem can be equivalently achieved by decomposing the corresponding specific instantiated problem, which is machine understandable and can be automatically conducted.

\textbf{Decomposition}. Finally, with instantiated network elements in $\mathcal{V}$, the dual function~(\ref{eq:dual}) of the network control problem $\mathbf{P}(\mathcal{V})$ can be obtained and then decomposed as described in Section~\ref{sec:theory}. To enable automated decomposition, the resulting dual function is represented using a tree. As illustrated in Fig.~\ref{fig:tree}, the whole dual function $\mathbf{P}$ is represented using the root node, which can be represented as the sum of a leaf node and an intermediate node, which can be further represented in a similar way. In this way, the decomposition of a network control problem (the dual function if dual decomposition is used) can be conducted in an automated fashion by traveling through all leaf nodes of the tree. The output of automated decomposition is a set of distributed subproblems each involving a single protocol layer and single network node. For each subproblem, a solution algorithm will be automatically generated and the resulting optimized network protocol parameters will be used to control the programmable protocol stack (PPS), which will be discussed in Section~\ref{sec:testbed}: WNOS Prototyping.


\subsection{Toy Example of DI-based Decomposition}\label{sub:toy}

Consider the following cross-layer network control problem:
\begin{eqnarray}\label{eq:genNUM}
\begin{array}{cl}
\mathrm{maximize}& \sum\limits_{s\in\mathcal{S}} R_s\\
\mathrm{subject\;to:}& \sum\limits_{s\in \mathcal{S}_l} R_s \leq C_l(\mathbf{\mathbf{\Pi}}),\;\forall l\in\mathcal{L}
\end{array}
\end{eqnarray}
\noindent where the objective
is to maximize the sum of rate $R_s$ of all flows $s\in\mathcal{S}$ at the transport layer; subject to the constraints that, for each link $l\in\mathcal{L}$, the aggregate rate of all the flows in $\mathcal{S}_l$, i.e., the set of links sharing link $l$, cannot exceed the capacity of the link $C_l(\mathbf{\mathbf{\Pi}})$ achievable with transmission strategies $\mathbf{\Pi}$ at the physical layer; by jointly controlling $R_s$ and $\mathbf{\Pi}$. Next, we show how the problem can be decomposed into two single-layer control problem through DI-based decomposition, while more examples of the DI-based decomposition  that consider different network problems will be discussed in Section~\ref{sec:evaluation}.


\textbf{Instantiation.}
As defined in Section~\ref{sec:decomp},  $\mathcal{S}$ (i.e., the set of all flows) and $\mathcal{L}$ (i.e., the set of all links) are global virtual elements while $\mathcal{S}_l \subset \mathcal{S}$ is a local virtual element. In favor of easy illustration, consider cardinality $N^{\mathrm{glb}} = 3$ for global virtual elements $\mathcal{S}$ and $\mathcal{L}$ and $N^{\mathrm{lcl}} = 2$ for local virtual elements $\mathcal{S}_l, \; \forall l\in\mathcal{L}$. Then, the global virtual elements $\mathcal{S}$ and $\mathcal{L}$ can be instantiated as  $\mathcal{S} = \{1, 2, 3\}$ (i.e., the network has in total three flows) and $\mathcal{L} = \{1, 2, 3\}$ (i.e., the network has in total three links). The instance of $\mathcal{S}$ will then be used as the mother set for instantiating local virtual elements $\mathcal{S}_l, \; \forall l\in\mathcal{L}$, as follows.

First, according to rule 1, i.e., equal cardinality, all $\mathcal{S}_l$ must have the same number of members.  
According to rule 2, no two or more $\mathcal{S}_l$ will be the same in the sense of ordered uniqueness.
If local virtual element $\mathcal{S}_l$, i.e., the set of sessions sharing link $l$, is instantiated to
$\{1, 2\} $ and $ \{2, 1\}$ for links $l=1$ and $l=2$, respectively, the resulting two instances will have the same set of ordered members, which violates rule 2 and hence are not allowed in DI.
An example instantiation that meets the two rules, which can be generated by a combination of peer randomly sampling and hash checking as discussed earlier in this section, is $\mathcal{S}_1 = \{1, 2\}$, $\mathcal{S}_2 = \{1, 3\}$ and $\mathcal{S}_3 = \{2, 3\}$. Let $\mathcal{L}_{s}$ represent the set of links used by flow $s$. Then, according to the instances for $\mathcal{S}_l$, the instances for $\mathcal{L}_{s}\subset \mathcal{L}$ can be given as $\mathcal{L}_{1} = \{1, 2\}$, $\mathcal{L}_{2} = \{1, 3\}$ and $\mathcal{L}_{3} = \{2, 3\}$. As a result, problem (\ref{eq:genNUM}) can be instantiated as
\begin{eqnarray}\label{eq:genNUMinst}
\begin{array}{cl}
\mathrm{maximize}& R_1 + R_2 + R_3\\
\mathrm{subject\;to:}& R_1 + R_2 \leq C_1(\mathbf{\Pi})\\
& R_1 + R_3 \leq C_2(\mathbf{\Pi})\\
& R_2 + R_3 \leq C_3(\mathbf{\Pi})
\end{array}.
\end{eqnarray}\vspace{-2mm}

\textbf{Decomposition.} Consider dual decomposition as discussed in Section~\ref{sec:theory}, then the  dual function of (\ref{eq:genNUMinst}) can be written as
\begin{align}\label{eq:instDual}
\mathrm{maximize}\; & R_1 + R_2 + R_3 + \lambda_1 [C_1(\mathbf{\Pi}) - R_1 - R_2] \nonumber\\
  & + \lambda_2 [C_2(\mathbf{\Pi}) - R_1 - R_3] + \lambda_3 [C_3(\mathbf{\Pi}) - R_2 - R_3],
\end{align}
where $\lambda_1, \lambda_2$ and $ \lambda_3$ are dual coefficients. By decomposing (\ref{eq:instDual}), problem (\ref{eq:genNUMinst}) can be decomposed into two single-layer problems:
\begin{eqnarray}
&\mathrm{Transport\; Layer}: \left \{ \begin{array}{l}
\mathrm{maximize}\;  R_1 - \lambda_1 R_1 - \lambda_2 R_1, \; \mathrm{for}\; s=1\\
\mathrm{maximize}\;  R_2 - \lambda_1 R_2 - \lambda_3 R_2, \; \mathrm{for}\; s=2\\
\mathrm{maximize}\;  R_3 - \lambda_2 R_3 - \lambda_3 R_3, \; \mathrm{for}\; s=3
\end{array}\right. \\
& \mathrm{Physical\; Layer}:\; \mathrm{maximize}\;  \lambda_1C_1(\mathbf{\Pi}) + \lambda_2C_2(\mathbf{\Pi}) + \lambda_3C_3(\mathbf{\Pi}),
\end{eqnarray}
where, at the transport layer, each flow $s\in \{1,2,3\}$ maximizes its own utility by adjusting its transmission rate $R_s$ with given dual coefficients; while the physical-layer subproblem maximizes a weighted-sum-capacity by adapting the transmission strategies $\mathbf{\Pi}$.

\textbf{Application to Run-time.} We show how the the decomposition results can be applied at network run time by
taking the transport-layer subproblem for $s=1$ as an example while the same principles can also be applied to other subproblems. For $s=1$, the utility of the subproblem can be rewritten as $R_1 - (\lambda_1 + \lambda_2) R_1$. Then, to determine the dual coefficients of $R_1$ at run time, we only need to identity the local virtual element corresponding to instance $\{\lambda_1, \lambda_2\}$, which is virtual element $\mathcal{L}_1$ according to the instantiation results of $\mathcal{L}_s$. This means that, at run time, the dual coefficients for flow $s$ should be collected, e.g., at the source node of flow $s$, from those links used by the flow.

\begin{figure*}[t]
\centering
\includegraphics[width=0.85\textwidth]{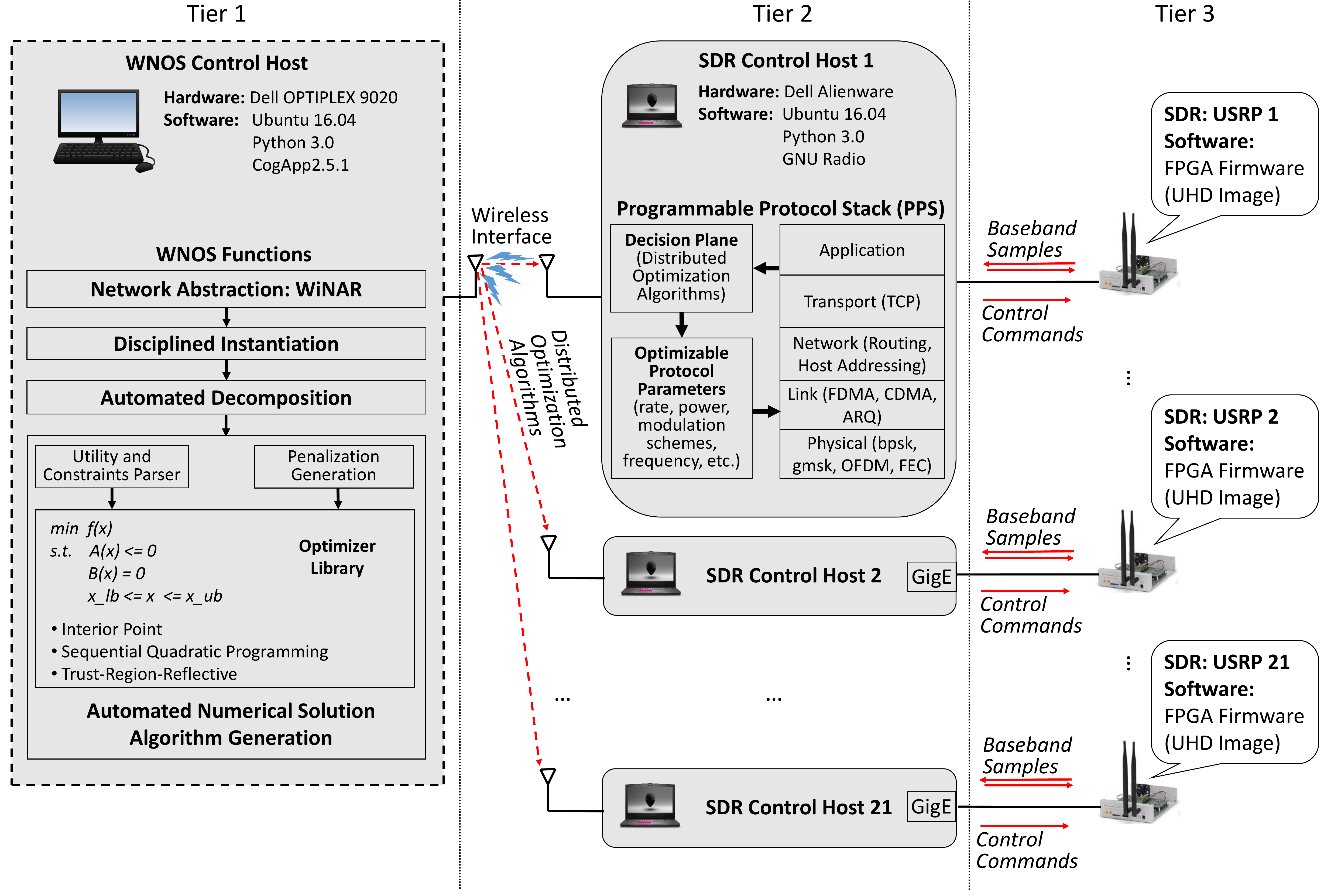} \caption{\small Prototyping diagram of WNOS.}
\label{fig:prototype}
\end{figure*}

\section{WNOS Prototyping} \label{sec:testbed}
So far, we have described the basic design principles of network abstraction and automated network control problem decomposition. To validate the proposed new ideas, we prototyped WNOS over a testbed with software defined radios.
This is however not easy because of several challenges: (i) with WNOS, one should be able to deploy a large scale  network by creating only a single piece of code to define the network control objective in a centralized manner. Since different SDR front-ends are controlled by different hosts, it is challenging to distribute and synchronize the generated code among the hosts; and (ii) there is no existing programmable protocol stack (PPS) that supports cross-layer control with optimizable protocol parameters at each layer. To address these challenges, next we first discuss the prototyping approach and then describe the newly designed PPS.

\subsection{Prototyping Approach}\label{sub:protappr}
A proof of concept of WNOS has been deployed over a network with 21 USRP software radios. The prototyping diagram is illustrated in Fig.~\ref{fig:prototype}, which follows a hierarchical architecture with three tiers, i.e., WNOS control host, SDR control host and SDR front-end.

\textbf{Prototype Architecture}. At the top tier of the hierarchical architecture is the WNOS control host, based on which one can specify the network control objective using the provided network abstract framework WiNAR. The output of this tier is a set of automatically generated distributed solution algorithms, which will be sent to each of the SDR control hosts. At the second tier, the programmable protocol stack (PPS) is installed on each of the SDR control hosts. The distributed optimization algorithms received from the WNOS control host are stored at the decision plane of the PPS. At run time,  the PPS will be compiled to generate operational code to control the SDR front-ends of the third tier. Finally, each of the SDR front-ends (i.e., USRP) receives the baseband samples from its control host via Gigabit Ethernet (GigE) interface and then sends them over the air with transmission parameters dynamically specified in the control commands from the SDR control hosts.

The primary benefit of prototyping WNOS based on an hierarchical architecture is to enable scalable network deployment. Specifically, the tier-1 WNOS control host is connected to all tier-2 SDR control hosts via wireless interfaces (which is Wi-Fi in current prototype), through which the generated distributed algorithms can be automatically \emph{pushed} to and installed at each of the SDR control hosts. Hence, one needs to create a single piece of code only in order to control all the 21 USRPs.

\textbf{WNOS Control Host}. On the WNOS control host, which is a Dell OPTIPLEX 9020 desktop running Ubuntu 16.04, four key WNOS functions have been implemented using a combination of Python 3.0 and CogApp 2.5.1, including the wireless network abstraction framework WiNAR, disciplined instantiation, automated decomposition as well as automated numerical solution algorithm generation (refer to Sections~\ref{sec:abst} and \ref{sec:autodecomp} for the techniques). We base our development on Python to take advantage of its high programming efficiency and high-level expressiveness \cite{python} and the flexible, open-source programming interfaces to GNU Radio for controlling USRPs. CogApp is an open-source software written in Python for template programming \cite{cogapp}, a programming technique based on which the automated numerical solution algorithm generation has been implemented in the current prototype.



\subsection{Programmable Protocol Stack}
As shown in Fig.~\ref{fig:prototype}, the programmable protocol stack (PPS) is installed on each of the five SDR control hosts, which are Dell Alienware running Ubuntu 16.04. The PPS has been developed in Python on top of GNU Radio to provide seamless controls of USRPs based on WNOS. To this end, a decision plane has been designed to install those distributed optimization algorithms generated by the WNOS control host and then \emph{pushed} to the SDR control hosts.

The developed PPS covers all the protocol layers. Based on the protocol stack, a multi-hop wireless ad hoc network testbed has been established using software-defined radio devices to verify the effectiveness of the designed wireless network operating system (WNOS).

%

\textbf{Application Layer}.
The application layer opens end-to-end sessions for transferring custom data such as files, binary blobs, as well as random generated data, among others.
A session can be established between any two network entities and multiple sessions can be established at the same time. Programmable parameters include the number of sessions and the number of hops in each session, as well as
the desired behavior of each session, e.g., maximum/minimum rate, power budget of the nodes, among others.

\textbf{Transport Layer}. The transport layer implements segmentation, flow control, congestion control as well as addressing.
This layer supports end-to-end, connection-oriented and reliable data transfer. To accomplish this, a \emph{Go-Back-N} sliding window protocol is implemented for flow control and congestion control, and
transport layer acknowledgments are used to estimate the end-to-end Round Trip Time (RTT), which serves as an estimate of network congestion. Programmable parameters at this layer include transmission rate, sliding window size and
packet size, among others.


\textbf{Network Layer}.
This layer 
implements host addressing and identification, as well as packet routing. The network layer is not only agnostic to data structures at the transport layer, but it also does not distinguish between operations of the various transport layer protocols. Routing strategies can be programmed at this layer.

\textbf{Datalink Layer}.
The core functionalities of this layer include fragmentation/defragmentation, encapsulation, network to physical address translation, padding, reliable point-to-point frame delivery, Logical Link Control (LLC) and Medium Access Control (MAC) among others.
In particular, the reliable frame delivery employs an hybrid LLC's \emph{Stop and Wait} ARQ protocol and Forward Error Correction (FEC) mechanism (\emph{Reed-Solomon} coding), such that frames are padded with FEC code and retransmissions are performed when the link is too noisy. The FEC is dynamic, reprogrammable, and can automatically adapt to the wireless link conditions at fine granularity, by increasing or decreasing the channel coding rate based on the observed packet error rate. Programmable parameters at this layer include channel coding rate, maximum retransmission times, and target residual link-level packet error rate, among others.

\textbf{Physical Layer}.
The physical layer features both CDMA and OFDM access schemes, yet with a wide set of modulation schemes supported, including Binary phase-shift keying (BPSK), Quadrature phase-shift keying (QPSK), Gaussian Minimum Shift Keying (GMSK) among others.
Programmable parameters at the physical layer include modulation schemes, transmission power, and receiver gain, among others.

\section{Experimental Evaluation} \label{sec:evaluation}
We evaluate the effectiveness, flexibility as well as scalability of the proposed WNOS by conducting experiments on the developed WNOS prototype, which is a testbed on large-scale USRP testbed with 21 nodes. Next, we first demonstrate in Section~\ref{subsec:demodecomp} the automated network control problem decomposition by considering specific network, and then show the experimental evaluation results in \ref{sec:impl}.

\subsection{Automated Network Control Problem Decomposition} \label{subsec:demodecomp}


We showcase how WNOS works by taking the network control problem in \cite{JOPC05} as an example. 
The objective of the network control problem, referred to as JOCP in \cite{JOPC05}, is to maximize the sum utility of a set of concurrent sessions by jointly optimizing the transmission rate of each session at the transport layer and controlling the transmission power of each transmitter at the physical layer. The underlying mathematical model of the network control problem is given as
\begin{eqnarray}\label{eq:jocp}
\begin{array}{cl}
\mathrm{maximize}& \sum\limits_{s\in\mathcal{S}} U_s(x_s)\\
\mathrm{subject\;to:}& \sum\limits_{s:l\in L(s)} x_s \leq c_l(\mathbf{P}),\;\forall l\in\mathcal{L}\\
&\mathbf{x,\;P} \succeq 0
\end{array}
\end{eqnarray}\vspace{-4mm}

\noindent where $\mathbf{x} \triangleq {(x_s)}$, with $x_s$ representing the transmission rate of session $s\in \mathcal{S}$, $\mathbf{P} \triangleq (P_n)$ is the transmission power profile of all the involved network nodes, $c_l(\mathbf{P})$ is the achievable capacity of link $l\in\mathcal{L}$ on path $L(s)$, and $U_s(x_s)$ is the achievable utility of session~$s$. Readers are referred to \cite{JOPC05} for details of the network control problem.

\begin{figure}[t]
\centering
\includegraphics[width=0.33\textwidth]{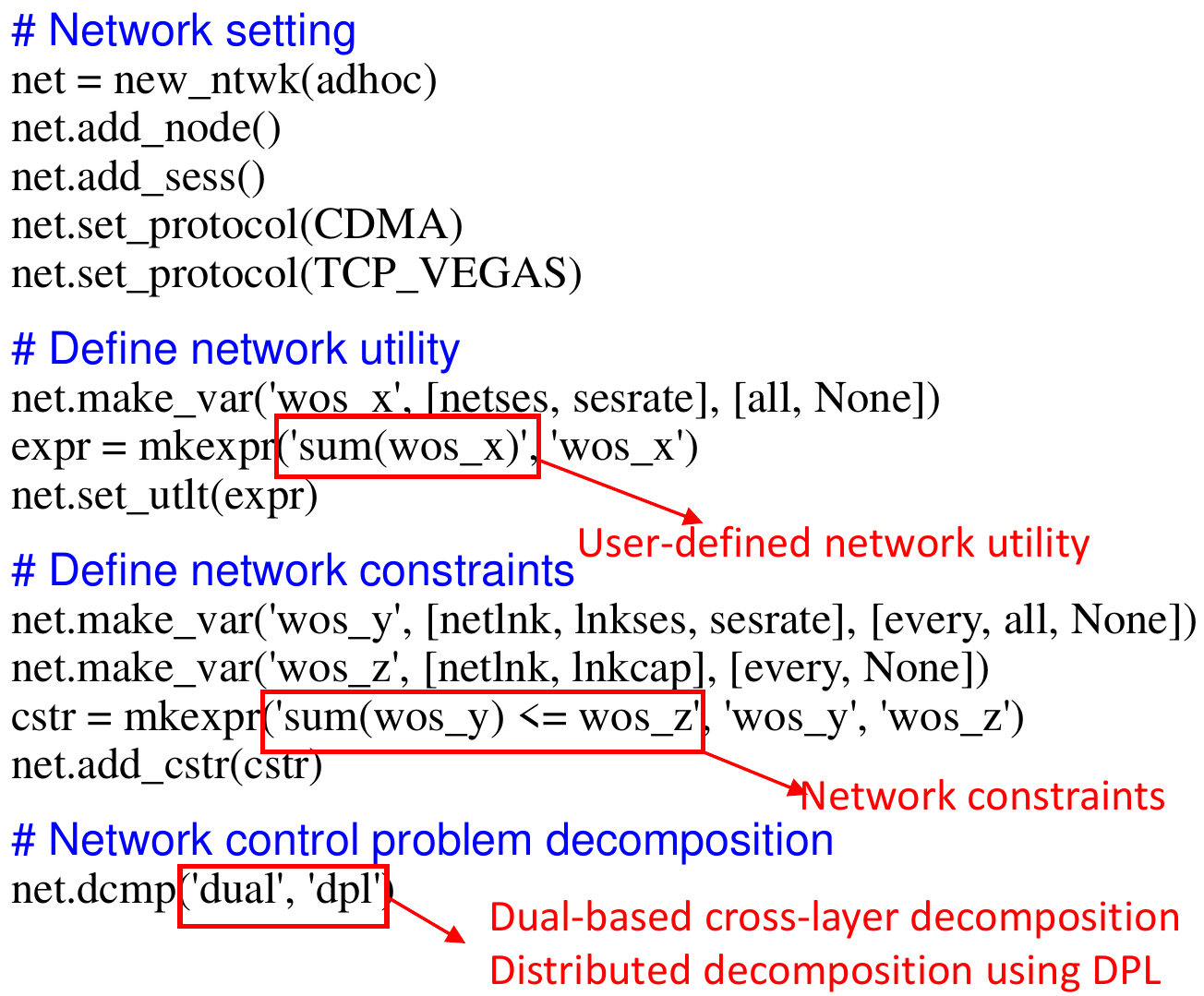} \vspace{-3mm}\caption{\small WNOS definition of network control problem based on the WiNAR. \vspace{-5mm}}
\label{fig:jocp}
\end{figure}

  \begin{figure}[b]
\centering
\includegraphics[width=0.45\textwidth]{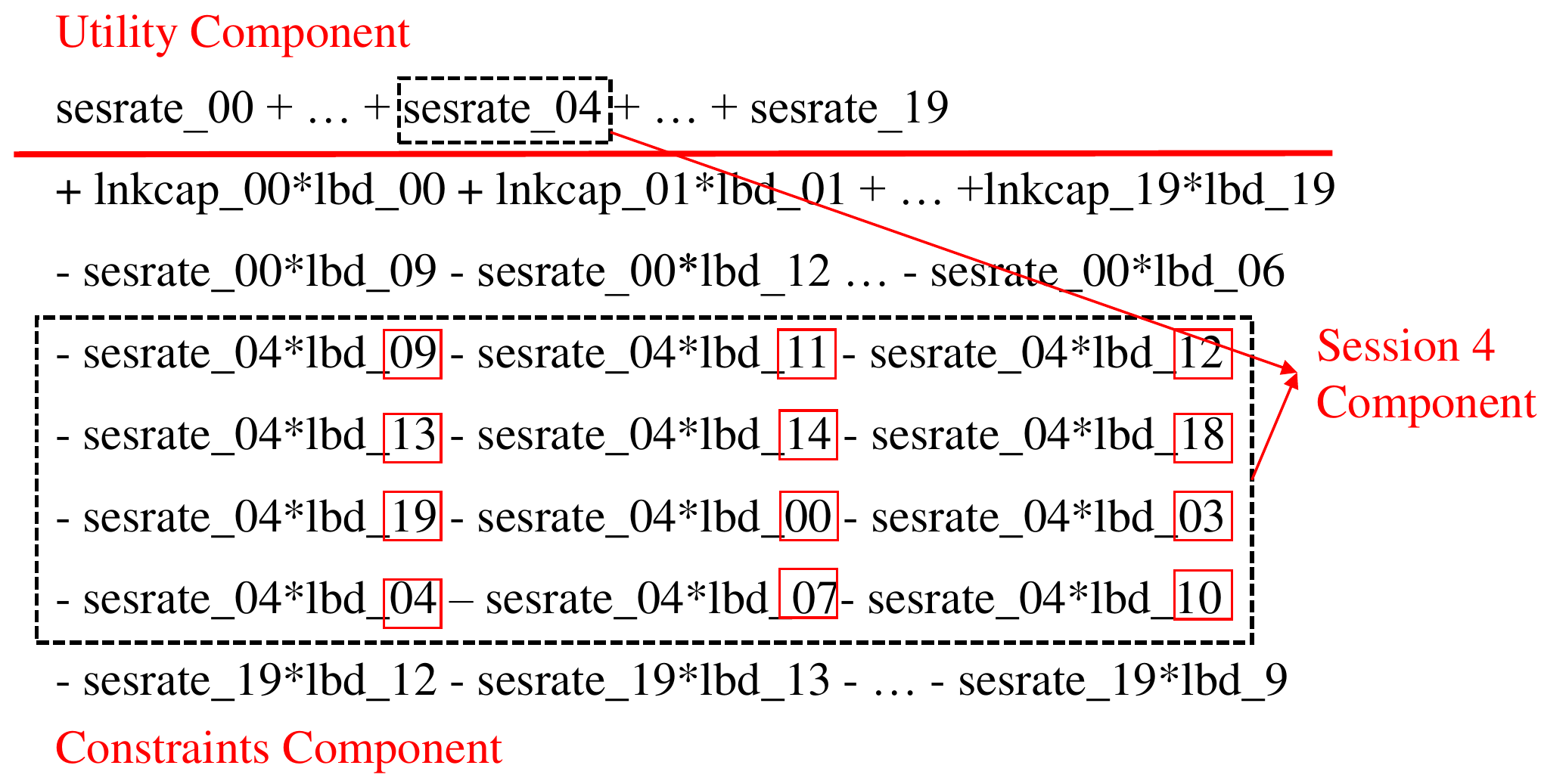} \vspace{-3mm}\caption{\small Dual function of the instantiated centralized network control problem.\vspace{-3mm} }
\label{fig:lambda}
\end{figure}

\textbf{WNOS Representation}. Based on the network abstraction interface provided by WNOS, i.e., \emph{WiNAR}, network control problem JOCP can then be defined in a high-level centralized abstract fashion as shown in Fig.~\ref{fig:jocp}, where network utility in (\ref{eq:jocp}) is defined as the sum rate of all sessions, i.e., $U_s(x_s) = x_s$ for each session $s\in\mathcal{S}$ in (\ref{eq:jocp}).
In the high-level definition in Fig.~\ref{fig:jocp}, there are three virtual network elements, i.e., netses, netlnk and lnkses, representing the set of all sessions, the set of all links and the set of links used by a session, respectively. 
The former two are \emph{global} elements representing the set of all sessions $\mathcal{S}$ and the set of all links $\mathcal{L}$ in the network, respectively. The third virtual element lnkses, i.e., $\{s:l\in L(s)\}$ in (\ref{eq:jocp}) represents the set of sessions sharing the same link, and hence is a \emph{local} element associated with each link instance of global virtual element netlnk. 

For instantiation of global virtual elements, the cardinality of the set of instances is by default set to 20, while it is set to 10 for local virtual element instantiation. Based on this, WNOS can generate up to 184756 unique instances for each abstract network element, which is sufficient to decompose moderate-size network control problems with up to hundreds of constraints.

Table~\ref{tbl:inst} shows the instantiation result of global virtual element $Links\;in\;Network$ and local virtual element $Sessions\; of\; Link$, where for each link instance a unique set of sessions sharing the link was constructed based on \emph{peer sampling} and \emph{hash checking} as described in Section~\ref{sec:decomp}. The set of links used by a session instance can then be instantiated accordingly, e.g., $\{0, 3, 4, 7, 9, 10, 11, 12, 13, 14, 18, 19\}$ for Session 4 as underlined in Table~\ref{tbl:inst}.

\small
 \begin{table}[b]
 \caption{\small Instantiation of virtual element $Sessions\; of\; Link$, i.e., lkses in Fig.~\ref{fig:jocp}, $s: l\in L(s)$ in (\ref{eq:jocp}).
 The set of all links is initiated to $\{1,\cdots,19\}$. Underlined links are links used by session 4.}
 \centering
 \begin{tabular}{c l}
 \hline\hline
 Link & Session Instances\\
 \hline
 \underline{0} & 3, \framebox{4}, 6, 7, 8, 14, 15, 17, 18, 19\\
 1 & 0, 2, 3, 6, 8, 10, 11, 12, 16, 17\\
 2 & 0, 5, 6, 7, 11, 13, 14, 15, 18, 19\\
 \underline{3} & 0, 1, \framebox{4}, 6, 10, 11, 13, 16, 17, 19\\
 \underline{4} & 0, 3, \framebox{4}, 7, 8, 12, 13, 14, 18, 19\\
 5 & 1, 3, 6, 10, 11, 12, 13, 15, 18, 19\\
 6 & 0, 1, 3, 5, 6, 7, 11, 14, 15, 16\\
 \underline{7} & 1, 3, \framebox{4}, 6, 12, 14, 15, 16, 17, 19\\
 8 & 1, 2, 5, 6, 7, 8, 9, 10, 12, 14\\
 \underline{9} & 0, 2, 3, \framebox{4}, 5, 12, 13, 15, 16, 17\\
 \underline{10} & 0, 1, \framebox{4}, 6, 7, 8, 9, 11, 16, 19\\
 \underline{11} & \framebox{4}, 5, 6, 7, 14, 15, 16, 17, 18, 19\\
 \underline{12} & 2, 3, \framebox{4}, 6, 7, 8, 12, 15, 17, 18\\
 \underline{13} & \framebox{4}, 6, 8, 13, 14, 15, 16, 17, 18, 19\\
 \underline{14} & 0, 1, 3, \framebox{4}, 5, 6, 8, 12, 16, 17\\
 15 & 2, 3, 6, 10, 11, 12, 13, 15, 16, 19\\
 16 & 1, 2, 3, 5, 6, 8, 9, 12, 15, 16\\
 17 & 1, 2, 7, 8, 12, 13, 14, 16, 18, 19\\
 \underline{18} & 0, 1, 3, \framebox{4}, 6, 7, 12, 13, 18, 19\\
 \underline{19} & 0, 2, \framebox{4}, 5, 8, 12, 14, 15, 16, 17\\
 \hline
 \end{tabular}
 \label{tbl:inst}
 \end{table}
 \normalsize

\textbf{Problem Decomposition}. Consider dual-based cross-layer decomposition (refer to Section~\ref{sec:theory} for the decomposition theory) as specified in the high-level abstract network control problem definition in Fig.~\ref{fig:jocp}. The resulting dual representation of the user-defined centralized network control (\ref{eq:jocp}) is given in Fig.~\ref{fig:lambda}, where the network constraints of (\ref{eq:jocp}) (constraints component) have been absorbed into the utility function (utility component), by introducing dual coefficients $lbd\_id$ with $id$ being the index of the link instance to which each dual coefficient is associated. Our objective is to decompose the initial user-defined centralized network control problem by decomposing the corresponding dual representation into a set of sub-problems each involving a single network element, e.g., a single session 4 as shown in Fig.~\ref{fig:lambda}.

\begin{figure}[t]
\centering
\includegraphics[width=0.45\textwidth]{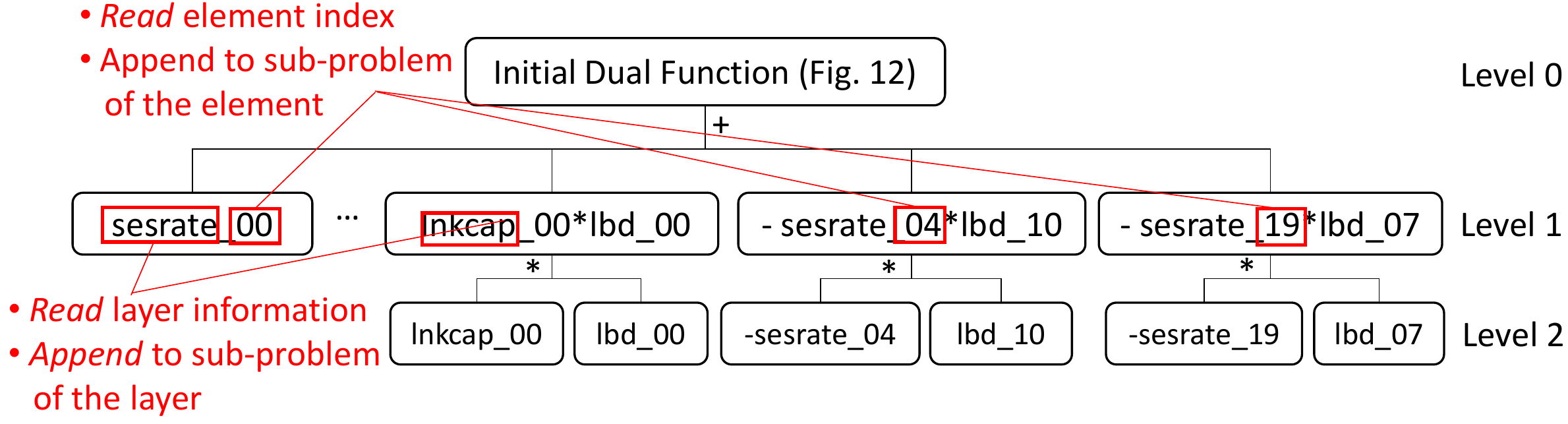} \vspace{-3mm}\caption{\small Tree representation of the instantiated dual function.\vspace{-3mm} }
\label{fig:dualtree}
\end{figure}

To this end, the dual representation is further represented as a three-level tree of sub-expressions, as shown in Fig.~\ref{fig:dualtree}, where level 0 is the initial dual expression in Fig.~\ref{fig:lambda}, level 1 comprises sub-expressions of sum operation in the initial representation, while each expression at level 1 can be further represented as a multiplication of two sub-expressions at level 2. Then, to decompose the user-defined central network control problem, we only need to walk over all level-1 elements of the tree and determine to which subproblem each of the elements should be categorized. For cross-layer decomposition, this can be accomplished as follows:
\begin{itemize}
\item For each level-1 sub-expression, extract the protocol layer information of the network element involved in the sub-expression using \emph{Read} operations defined in Section~\ref{sec:abst}.
\item Categorize the sub-expression into the sub-problem of the corresponding protocol layer.
\end{itemize}
The decomposition will result in a set of subproblems each involving only a single protocol layer. For example, in Fig.~\ref{fig:dualtree}, level-1 element $sesrate\_00$ will be categorized into the transport-layer subproblem because the network element $sesrate$ is a transport layer parameter representing the source rate of a session. Accordingly, subexpression $lnkcap\_00*lbd\_00$ will be categorized as a subproblem corresponding to the physical layer.

Similarly, each of the resulting sub-problems can be further decomposed into sub-problems each involving a single network element, e.g., node, session, so that they can be solved in a distributed fashion and result in distributed control actions. As shown in Fig.~\ref{fig:dualtree}, this can be accomplished as follows:
\begin{itemize}
\item Extract the index information of the network element involved in each level-1 sub-expression, e.g., 00 for level-1 subexpression $sesrate\_00$.
\item Categorize the sub-expression into the sub-problem corresponding to a distributed subproblem, e.g., categorize $sesrate\_00$, $-sesrate\_04*lbd\_10$ and $-sesrate\_19*lbd\_07$ as the subproblems corresponding to sessions 0, 4 and 19, respectively.
\end{itemize}

\begin{figure*}[t]
\begin{center}
\begin{tabular}{cc}
\hspace{-6mm}\includegraphics[width=6.15cm]{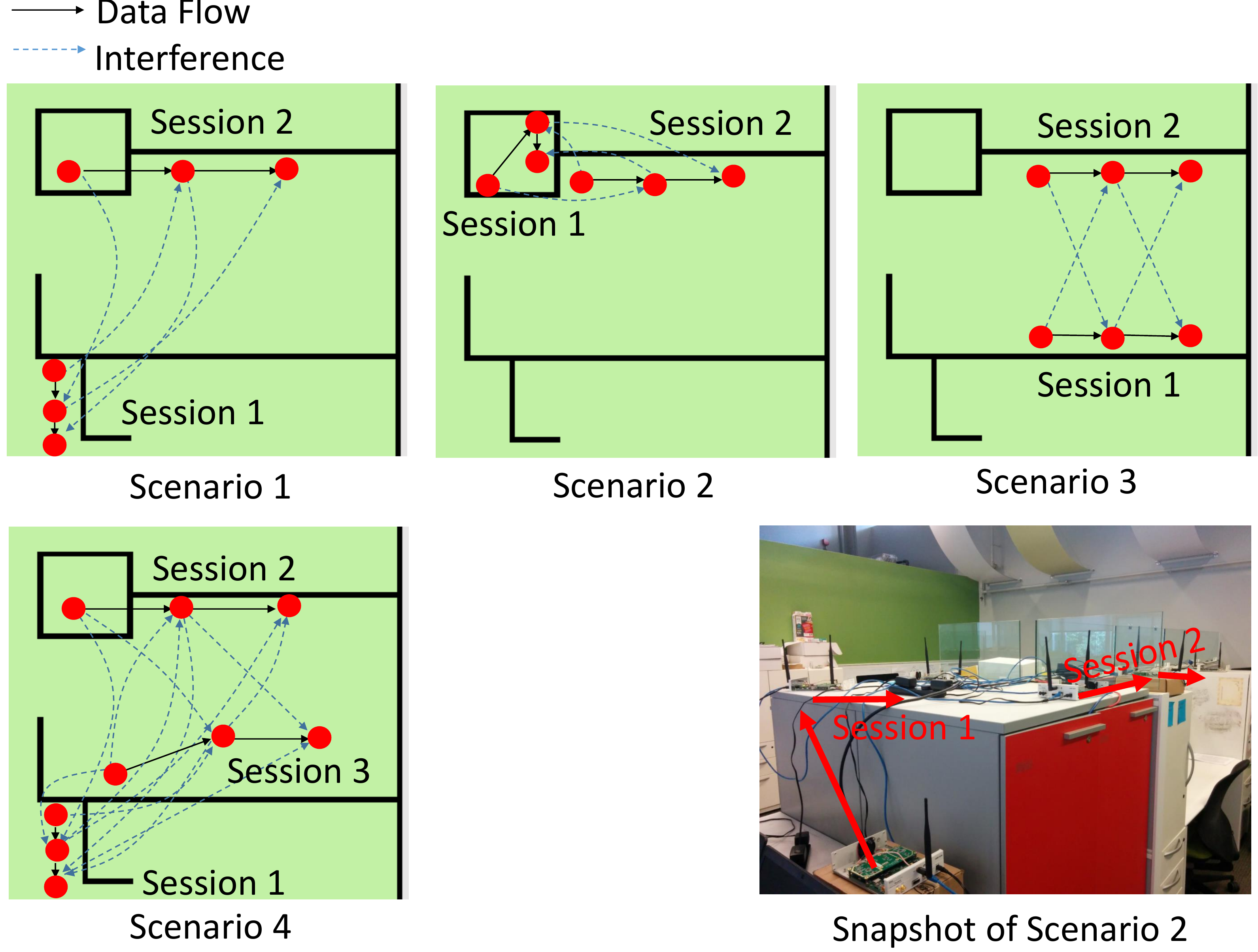}&
\hspace{-3mm}\includegraphics[width=12cm]{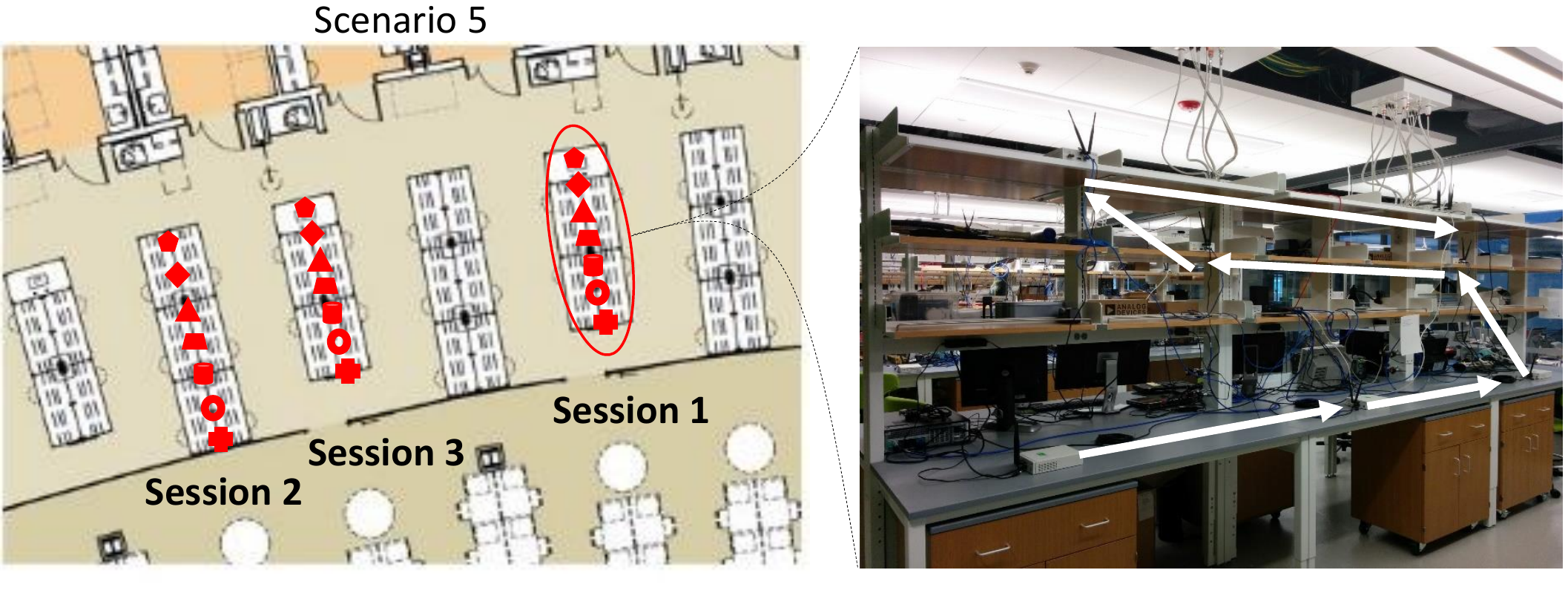}\\
\small (a) & \small (b) \vspace{-4mm}
\end{tabular}
\caption{\small  Experimental Scenarios: (a) Scenarios 1-4 and (b) Scenario 5.\vspace{-1mm}} \label{fig:testtopo}
\end{center}
\end{figure*}

\emph{Mapping From Instantiation to Abstract Domain}: Recall in Section~\ref{sec:decomp} that our objective is to construct a set of instantiations of the user-defined high-level abstract network control so that decomposing any of the problem instantiations decomposes the abstract problem. As described in Section~\ref{sec:decomp}, this is guaranteed by the one-to-one mapping between virtual network elements and their instantiations obtained through peer-sampling and hash checking. Next, we show the one-to-one mapping taking the following instantiated transport-layer subproblem as an example.
\vspace{-2mm}
\begin{eqnarray}\label{eq:examp}
\begin{array}{ll}
&sesrate\_04 - sesrate\_04*(lbd\_09 + lbd\_11 \\
&+ lbd\_12 + lbd\_13 + lbd\_14 + lbd\_18 + lbd\_19  \\
&+ lbd\_00  + lbd\_03 + lbd\_04 + lbd\_07 + lbd\_10).
\end{array}
\end{eqnarray}\vspace{-4mm}

\noindent We can see that (\ref{eq:examp}) is the subproblem by categorizing those Session 4 components in Fig~\ref{fig:jocp}. In (\ref{eq:examp}), dual coefficients $lbd$ are parameters that will be received by the source node of session 4 from links with indexes \{09, 11, 12, 13, 14, 18, 19, 00, 03, 04, 07, 10\}, which is namely the instantiation set for local virtual element $Links\;of\;Session$ for Session 4, as shown in Table~\ref{tbl:inst}. Hence, (\ref{eq:examp}) can be further represented for all sessions, which corresponds to the global virtual network element $Sessions\;of\; Network$,
\vspace{-2mm}
\begin{equation}\label{eq:examp2}
sesrate - sesrate*sum(\textbf{lbd})
\end{equation}\vspace{-6mm}

\noindent where $\textbf{lbd}$ represents the vector of dual parameters received by each source node of the session at network run time.



\textbf{Automated Solution Algorithm Generation}. For each of the subproblems resulting from automated network control problem decomposition described in Section~\ref{sec:decomp}, a numerical solution algorithm (e.g., interior-point method) is selected to solve the problem, and the optimization results are then fed into the decision plane of a Programmable Protocol Stack (PPS). The diagram of automated algorithm generation is illustrated in Fig.~\ref{fig:prototype}. Each of the subproblem instantiations, e.g., the transport layer subproblem in (\ref{eq:examp}), is fed into \emph{utility and constraints parser} and \emph{penalization generation} (which will be described below in this section) components to determine the required parameters by an \emph{algorithm template}, including optimization variable $x$, utility function $f(x)$ as well as inequality and equality constraint parameters $A$,  $B$, $x\_{lb}$ and $x\_{ub}$ in Fig.~\ref{fig:prototype}. Then, based on the requirements of the solution algorithms in terms of computing efficiency and optimality precision specified by network controllers via network control interface layer, a numerical solver is selected by the \emph{optimization engine} to obtain the globally optimal solution for convex problems and a suboptimal solution if the problems are nonconvex. A wide set of numerical solvers have been integrated into WNOS to meet diverse requirements in convergence speed and optimality precision for small, moderate and large-size network control problems, including Interior-Point (IP) method \cite{Byrd1999},  Sequential Quadratic Programming (SQP) \cite{Gill81} and Trust-Region-Reflective (TRR) \cite{Byrd2000}, among others. The output of the automated algorithm generator is a set of optimization algorithms that can be directly executed at each protocol layer to optimize a penalized version of local control objectives.

\textbf{Penalization}. As described in Section~\ref{sec:decomp}, the objective of penalization is to provide a signaling framework, based which the resulting distributed algorithms can exchange certain signaling messages in favor of improved network performance. In WNOS, this is accomplished by converting the mathematical expressions of each sub-problem obtained through decomposition into a symbolic domain. Take the physical-layer sub-problem as an example, where the objective is to maximize the sum capacity of all links with $\mathrm{lnkcap}$ expressed as
\vspace{-2mm}
\begin{equation} \label{eq:lnkcap}
lnkcap = freq*\log_2\left(1+\frac{lnkpwr * lnkgain}{lnknoise + lnkgain\_itf*itfpwr}\right)\vspace{-1mm}
\end{equation}

\noindent where $\mathrm{freq, lnkpwr, lnkgain, lnknoise, lnkitf, lnkgain\_itf}$ and $\mathrm{itfpwr}$ are parameters that can be measured online for each distributed subproblem.
Then, taking DPL as example in Section~\ref{sec:theory}, the penalization item can be obtained in an automated fashion in the symbolic domain by deriving the derivative of the $lnkcap$ expression with respect to the strategy of each interfering agent, i.e., $itfpwr$ in (\ref{eq:lnkcap}).

In (\ref{eq:lnkcap}) the link expression has been constructed for CDMA at the physical layer as specified in the high-level centralized abstract network control problem definition in Fig.~\ref{fig:jocp}. Based on WNOS, network designers are allowed to configure the network protocols operating at different layers, e.g., OFDM at physical layer,  and define their own network utilities. For example, in Fig.~\ref{fig:jocp} by defining the network utility as
$\mathrm{expr = mkexpr('sum(log(wos\_x))', 'wos\_x')}$, a new subproblem $\log(sesrate) - sesrate*sum(\textbf{lbd})$ will be obtained at the transport layer for each session,
which introduces proportional fairness among the sessions.

The generated solution algorithms, as discussed in Section~\ref{sec:arc}, are fed into the decision plane of the Programmable Protocol Stack (PPS), and the optimization results obtained by solving local network control problems are used to dynamically configure the protocol parameters for each SDR node. Next we evaluate the performance of WNOS on software-defined radio testbed in various networking scenarios.

\begin{figure*}[t]
\begin{center}
\begin{tabular}{cccc}
\hspace{-6mm}\includegraphics[width=4.5cm]{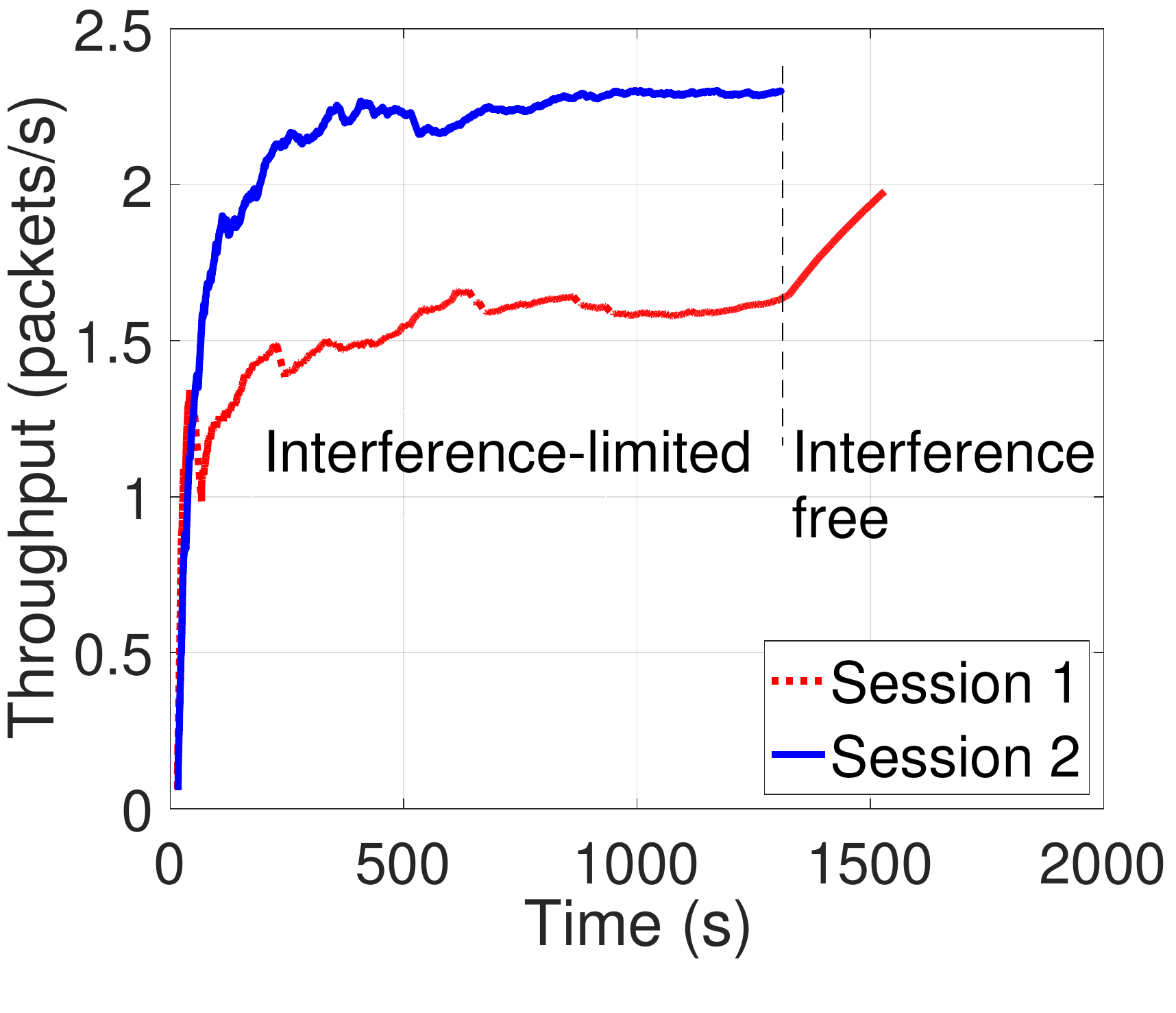}&
\hspace{-3mm}\includegraphics[width=4.6cm]{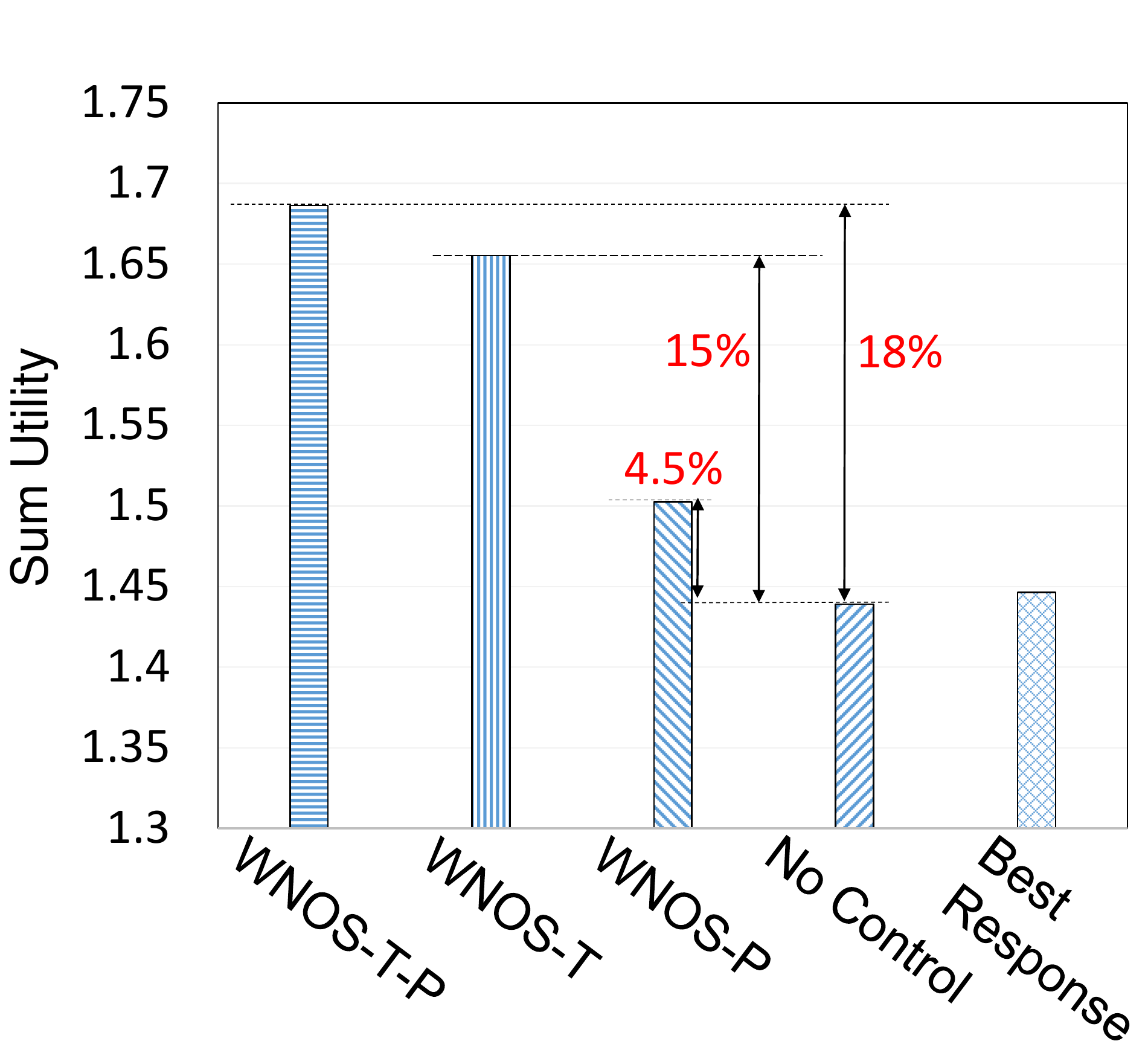}&
\hspace{-3mm}\includegraphics[width=4.6cm]{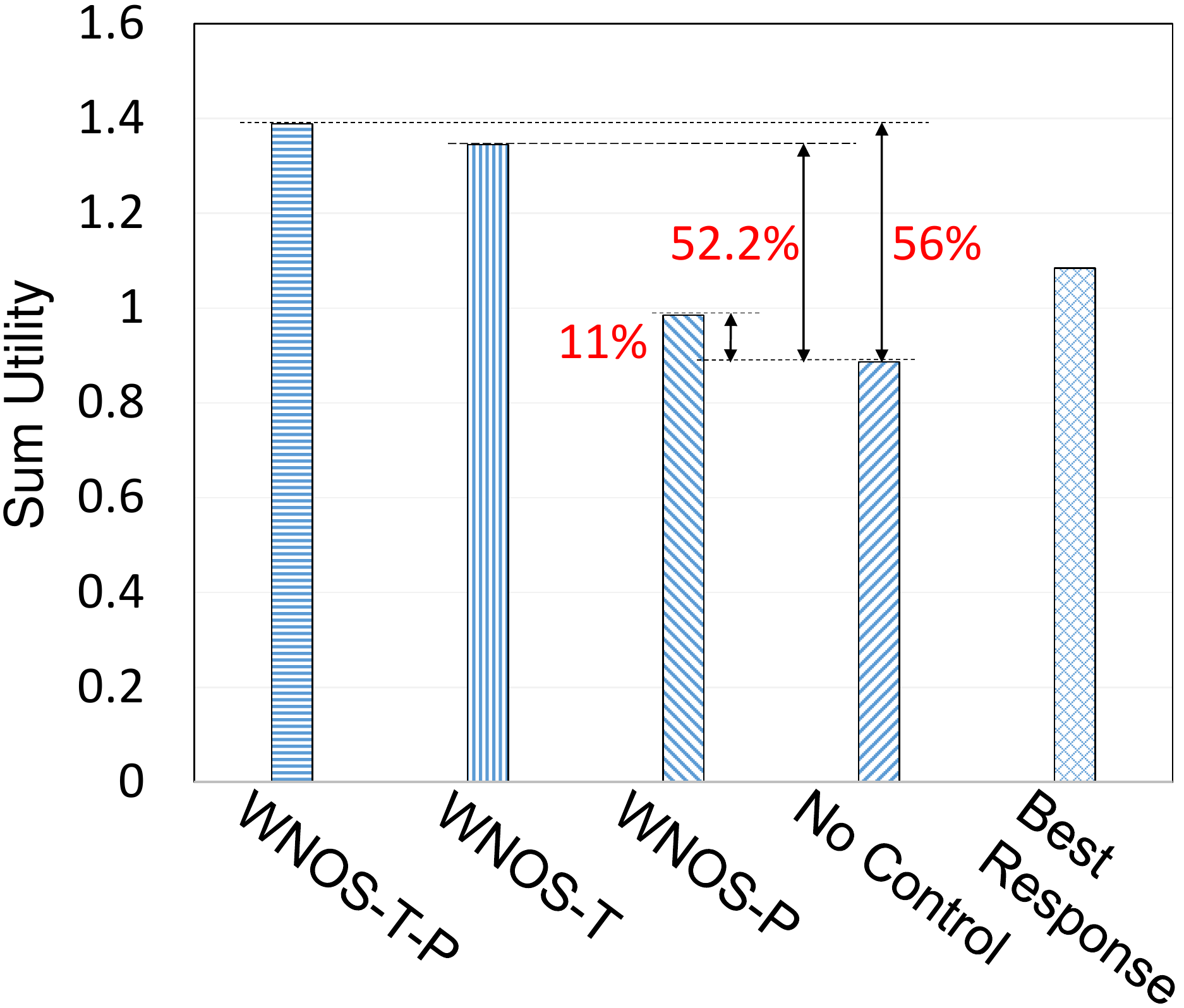}&
\hspace{-3mm}\includegraphics[width=4.6cm]{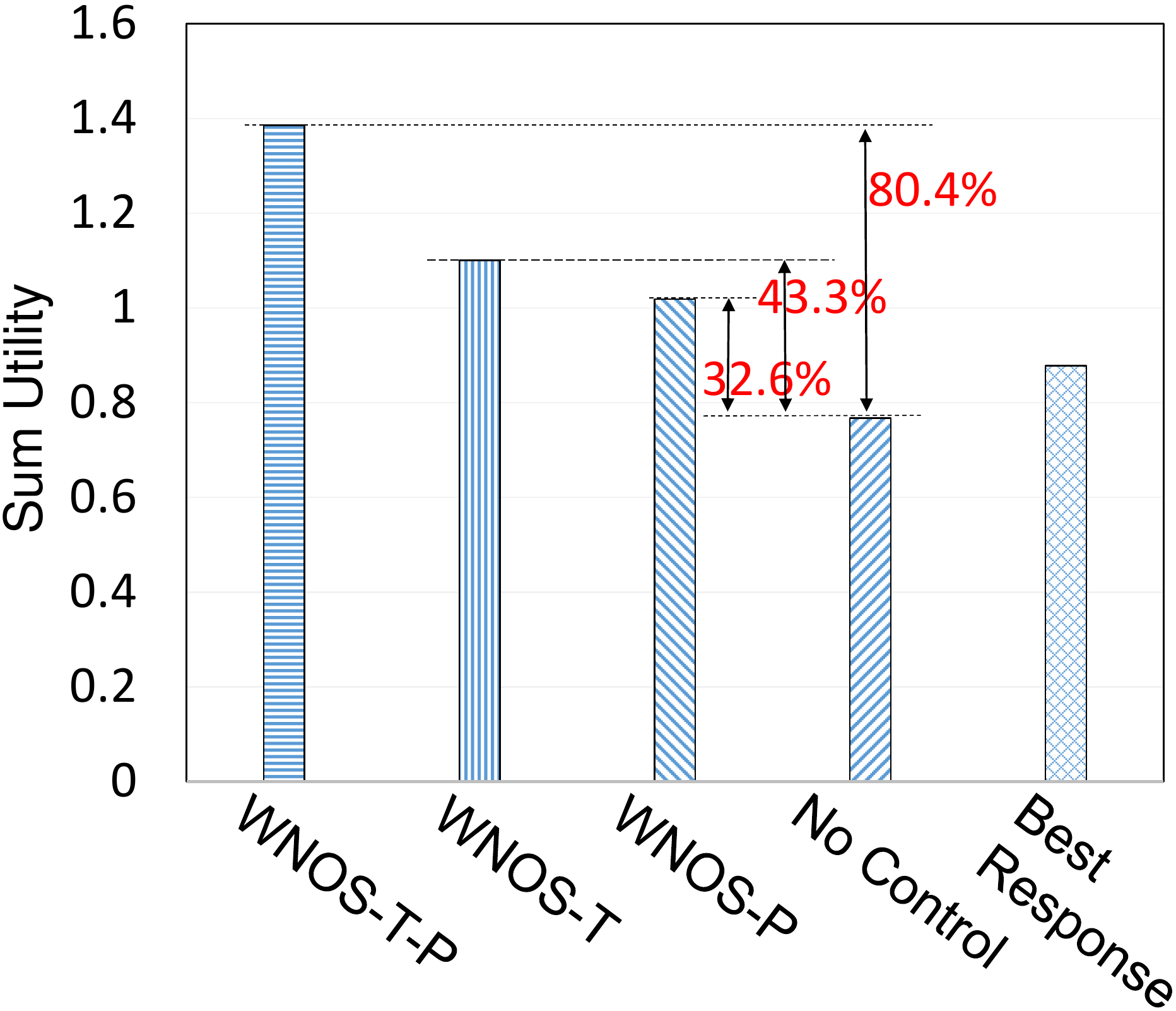}\\
\small (a) & \small (b) & \small (c) & \small (d)\vspace{-4mm}
\end{tabular}
\caption{\small  (a) End-to-end throughputs of sessions 1 and 2; Average sum utility of scenarios (b) 1, (c) 2 and (d) 3.\vspace{-1mm}} \label{fig:result}
\end{center}
\end{figure*}


%
%

\subsection{Software-defined Radio Implementation}\label{sec:impl}
We test WNOS on the designed SDR testbed in five different networking scenarios. As shown in Fig.~\ref{fig:testtopo}, Scenarios 1-3 deploy six nodes and two traffic sessions; while Scenario 4 considers nine nodes and three traffic sessions, with each session spanning over  two hops. In Scenario 5, three sessions are deployed over 21 nodes, with six hops for each session. Six spectrum bands in the ISM bands are shared by the 21 USRPs, with bandwidth of $200\;\mathrm{kHz}$ for each spectrum band. At each USRP, the data bits are first modulated using GMSK and then sampled at sampling rate of configured $800\;\mathrm{kHz}$. Reed-solomon (RS) code is used for forward error coding (FEC) with coding rate ranging from 0.1 to 0.4 at a step of 0.1.
The code to repeat experiments is available on website:
\url{http://www.ece.neu.edu/wineslab/WNOS.php}.


Through the experiments, we seek to demonstrate the following properties:
\begin{itemize}
\item {\bf Effectiveness}. Through experiments in Scenarios 1-3, we show that WNOS-based network optimization outperforms non-optimal or purely locally optimal (greedy) network control;
\item {\bf Flexibility}. Through experiments in Scenarios 4 and 5, we showcase  the flexibility of WNOS in modifying the global network behavior by changing control objectives and constraints.
\item {\bf Scalability}. In Scenario 5 we show the scalability of WNOS by deploying code over a large-scale network.
\end{itemize}

\textbf{Effectiveness}.
We show the effectiveness of WNOS on the developed SDR testbed.
At the physical layer, two spectrum bands are used, $1.3~\mathrm{GHz}$ and $2.0~\mathrm{GHz}$. If two transmitters (either source or relay) are tuned to the same spectrum band, their transmissions will interfere as shown in Fig.~\ref{fig:testtopo}(a). The control objective is to maximize the sum utility of the two sessions (referred to as Control Program 1, which can be specified by \emph{expr = mkexpr(`sum(log(wos\_x))', `wos\_x')} as in (\ref{eq:jocp})) by jointly
controlling the transmission rate at the transport layer and the transmission power at the physical layer. For each session, the utility is defined as the logarithm of the achievable end-to-end throughput. 

Five schemes have been tested: (i) WNOS-T-P: transport and physical layers are jointly controlled using the optimization algorithms automatically generated by WNOS; (ii) WNOS-T: only the transport layer rate is controlled by WNOS; (iii) WNOS-P: only the physical layer power is controlled by WNOS; (iv) neither transport or physical layer are controlled by WNOS; and (v) Best Response: maximum rate and power are used at the transport and physical layers, respectively. In all schemes, the initial operating points (i.e., rate and power) are randomly generated. Power control is implemented by controlling the transmit gain (which takes value from 0 to 30 $\mathrm{dB}$) of the FPGA of USRP N210s.



We first validate WNOS in an environment with tight coupling of different sessions via interference. Fig.~\ref{fig:result}(a) reports  the achievable end-to-end throughput vs time for the two sessions (in terms of $\mathrm{packets/s}$) in network scenario 2. The packet length is set to $2048\:\mathrm{bits}$. We observe that the throughput of the two sessions converges to $1.6$ and $2.3\:\mathrm{packets/s}$ when they are active simultaneously, i.e., in the interference-limited region in Fig.~\ref{fig:result}(a). After session 2 is done transmitting all of its packets (3000 packets), session 1 operates in the interference-free region and its throughput starts to increase significantly.

\begin{figure}[t]
\centering
\includegraphics[width=0.4\textwidth]{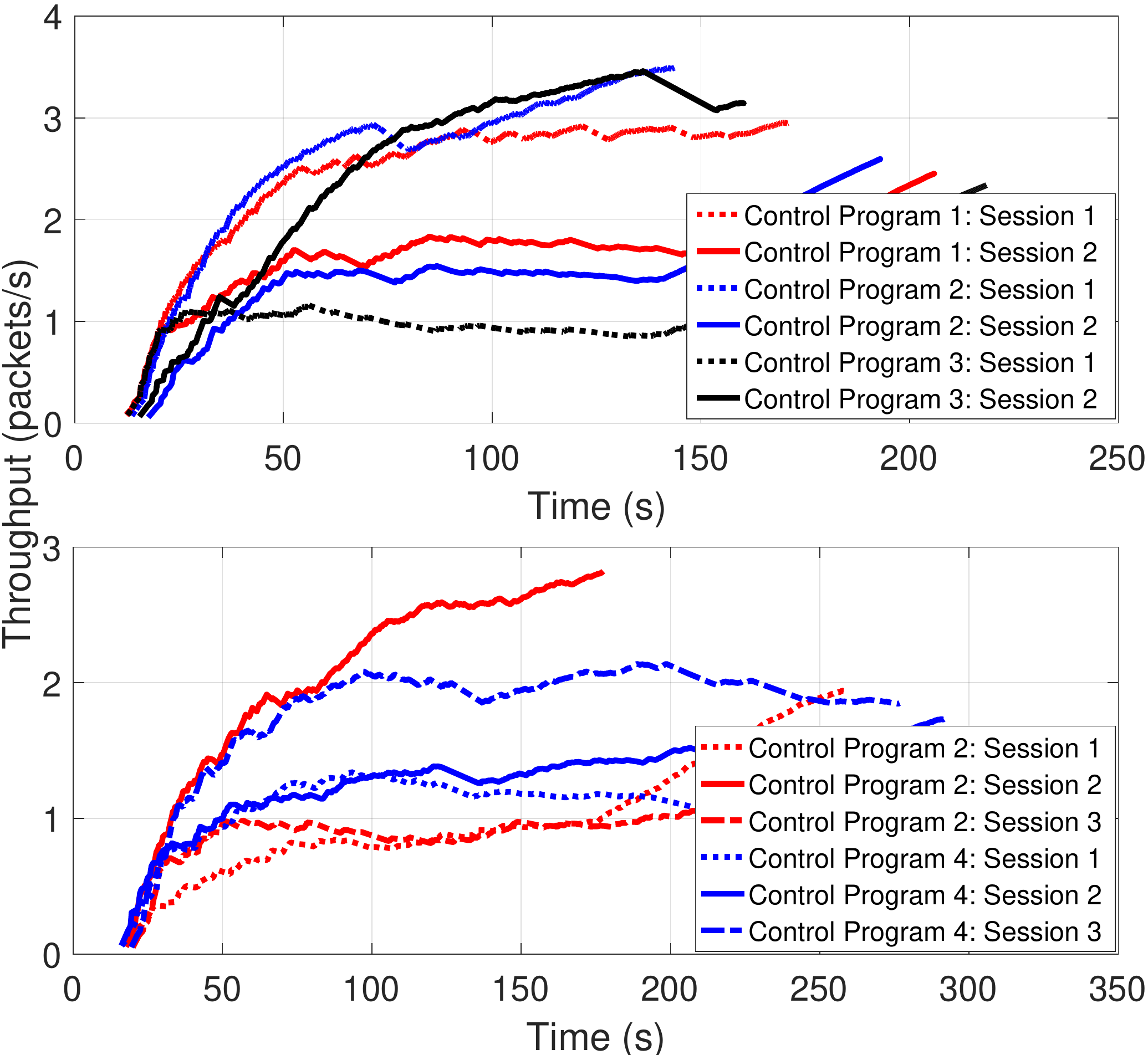} \vspace{-5mm}\caption{\small Network behaviors with different control programs. \vspace{-6mm}}
\label{fig:twoprob}
\end{figure}

\begin{figure*}[t]
\begin{center}
\begin{tabular}{cc}
\includegraphics[width=7.8cm]{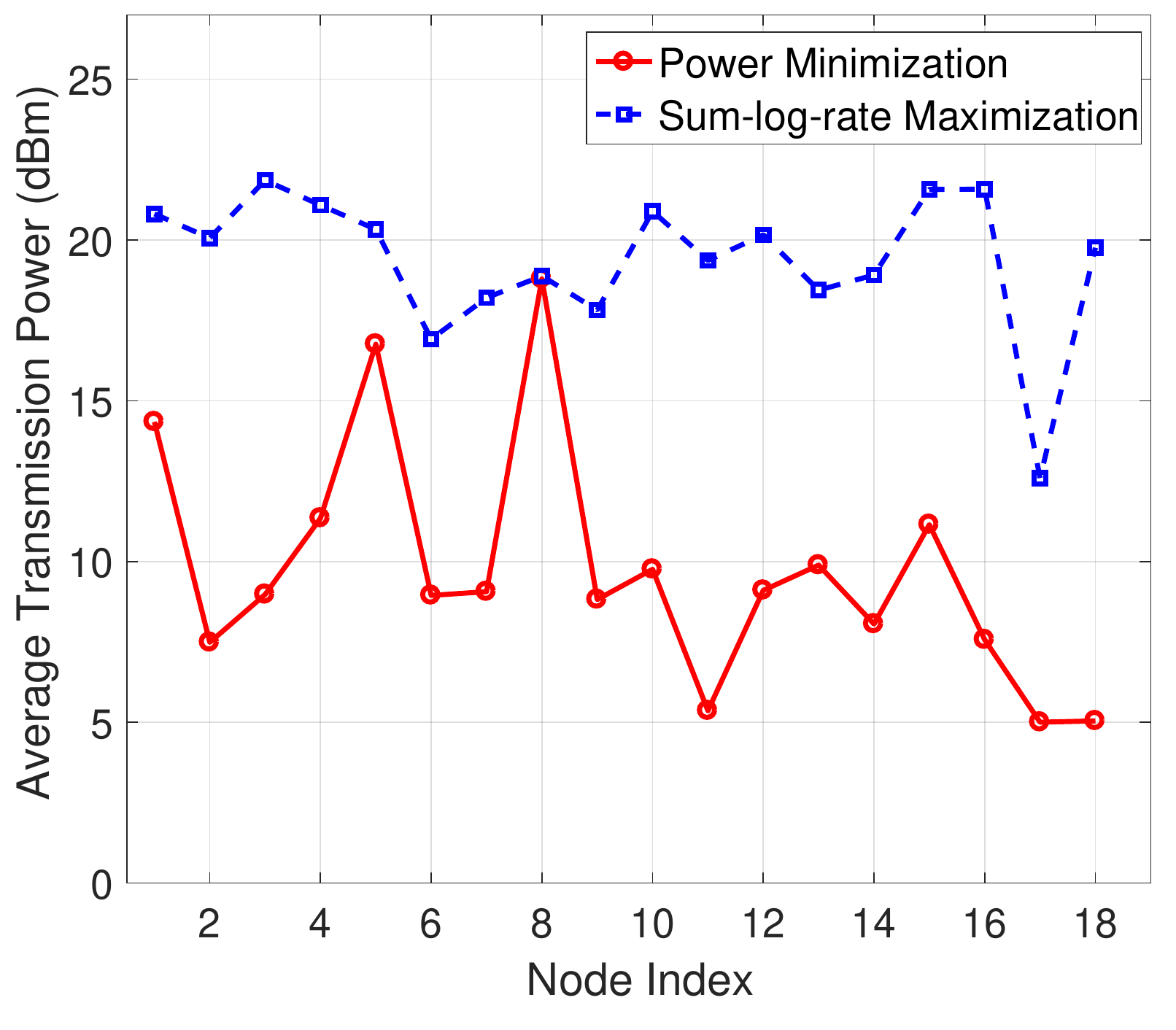}&
\includegraphics[width=8cm]{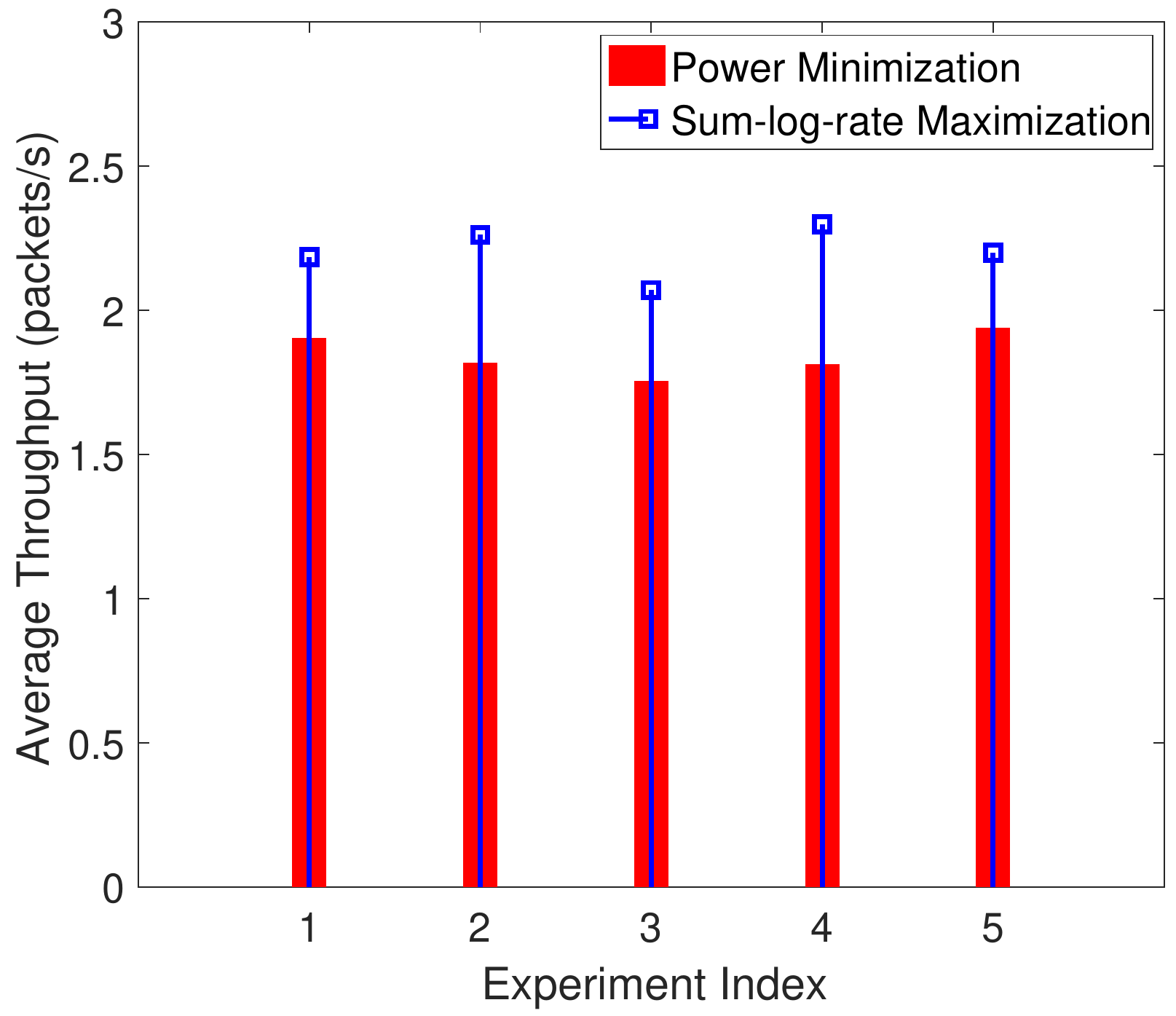}\\
\small (a) & \small (b) \vspace{-4mm}
\end{tabular}
\caption{\small (a) Transmission power and (b) throughput resulting from two different control objectives: sum-log-rate maximization and power minimization.\vspace{-1mm}} \label{fig:ISECAverage}
\end{center}
\end{figure*}

\begin{figure*}[t]
\begin{center}
\begin{tabular}{cc}
\includegraphics[width=8cm]{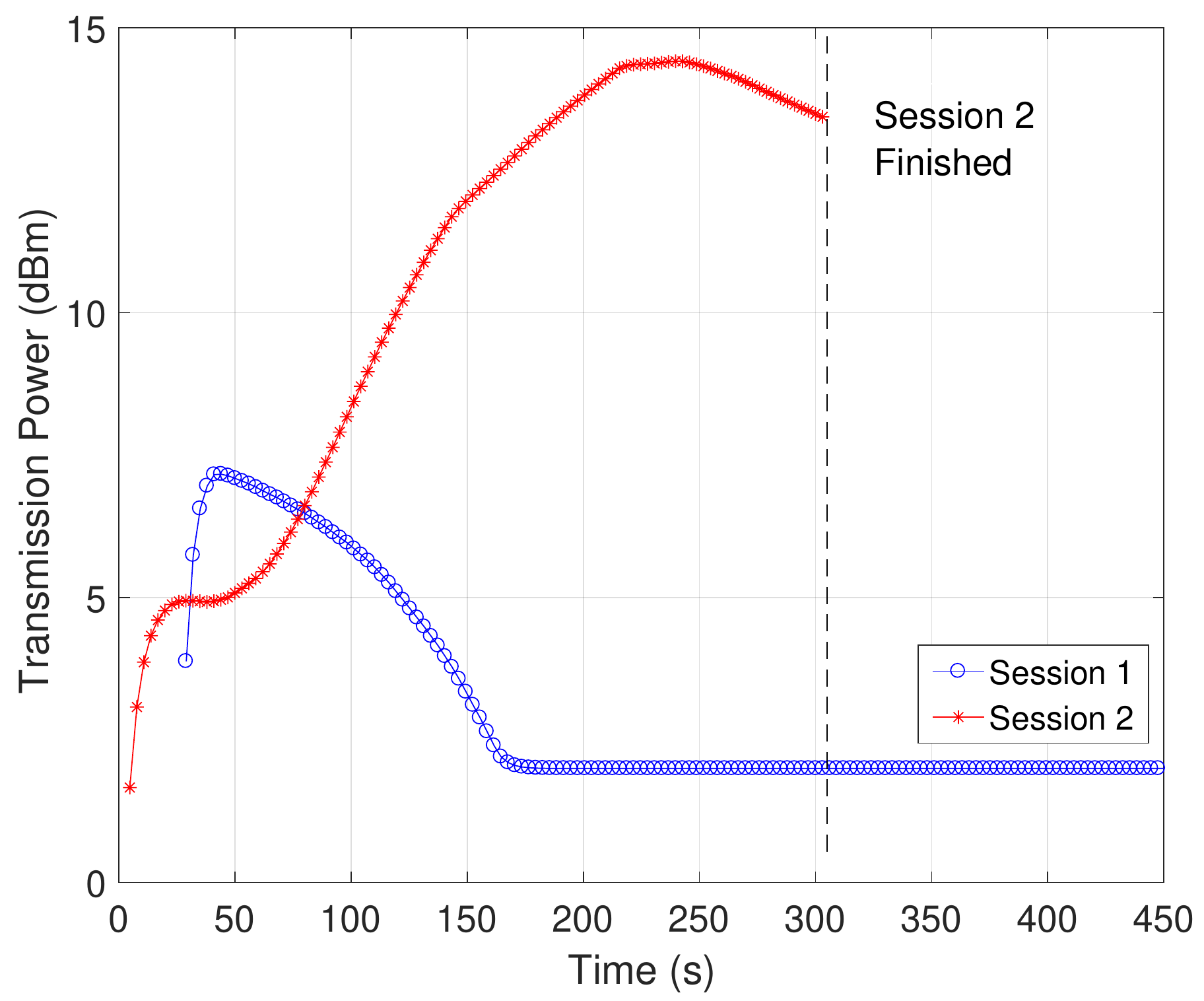}&
\includegraphics[width=8cm]{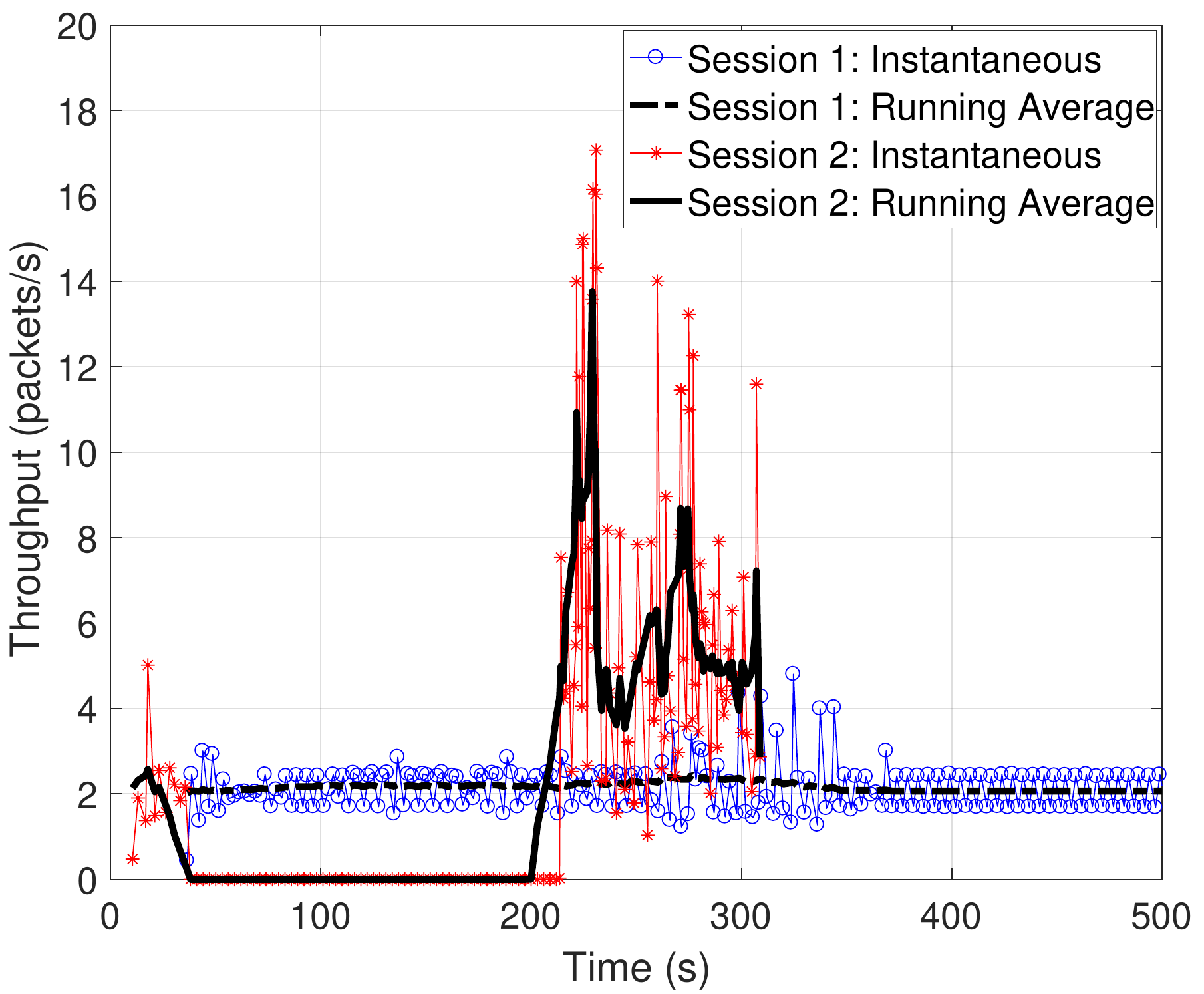}\\
\small (a) & \small (b) \vspace{-4mm}
\end{tabular}
\caption{\small Instance of (a) transmission power (source node) and (b) throughput resulting from power minimization.\vspace{-1mm}} \label{fig:pm}
\end{center}
\end{figure*}

The average performance of the five schemes is reported in Figs.~\ref{fig:result}(b), (c) and (d) for network scenarios 1, 2 and 3, respectively. As discussed above, the three network scenarios have been designed to present different levels of interference between the two sessions. The sum utility achievable by the best response scheme is a good indicator since with this scheme each node always transmits at the maximum power, i.e., $30~\mathrm{dB}$ transmit gain is used for USRP N210.  For example, with the least amount of interference in scenario 1, best response achieves the highest sum utility of 1.44 compared to 0.89 in network scenario 3. From the three figures, it can be seen that, compared with no control, considerable performance gain can be achieved by the WNOS-T-P, i.e., with transport and physical layers jointly controlled, and this gain increases as the interference level increases. Once more, we would like to emphasize that this is obtained by writing \emph{only a few lines of high level code on a centralized abstraction}; while the behavior is obtained through automatically generated distributed control programs. Specifically, up to 80.4\% utility gain can be achieved in network scenario 3, which has the highest interference. In the case of no cross-layer control, i.e., only one protocol layer is optimized, WNOS still achieves significant utility gain, which varies from 4.5\% to 52.2\% in the tested instances.

\begin{figure*}[t]
\begin{center}
\begin{tabular}{cc}
\hspace{-6mm}\includegraphics[width=8.1cm]{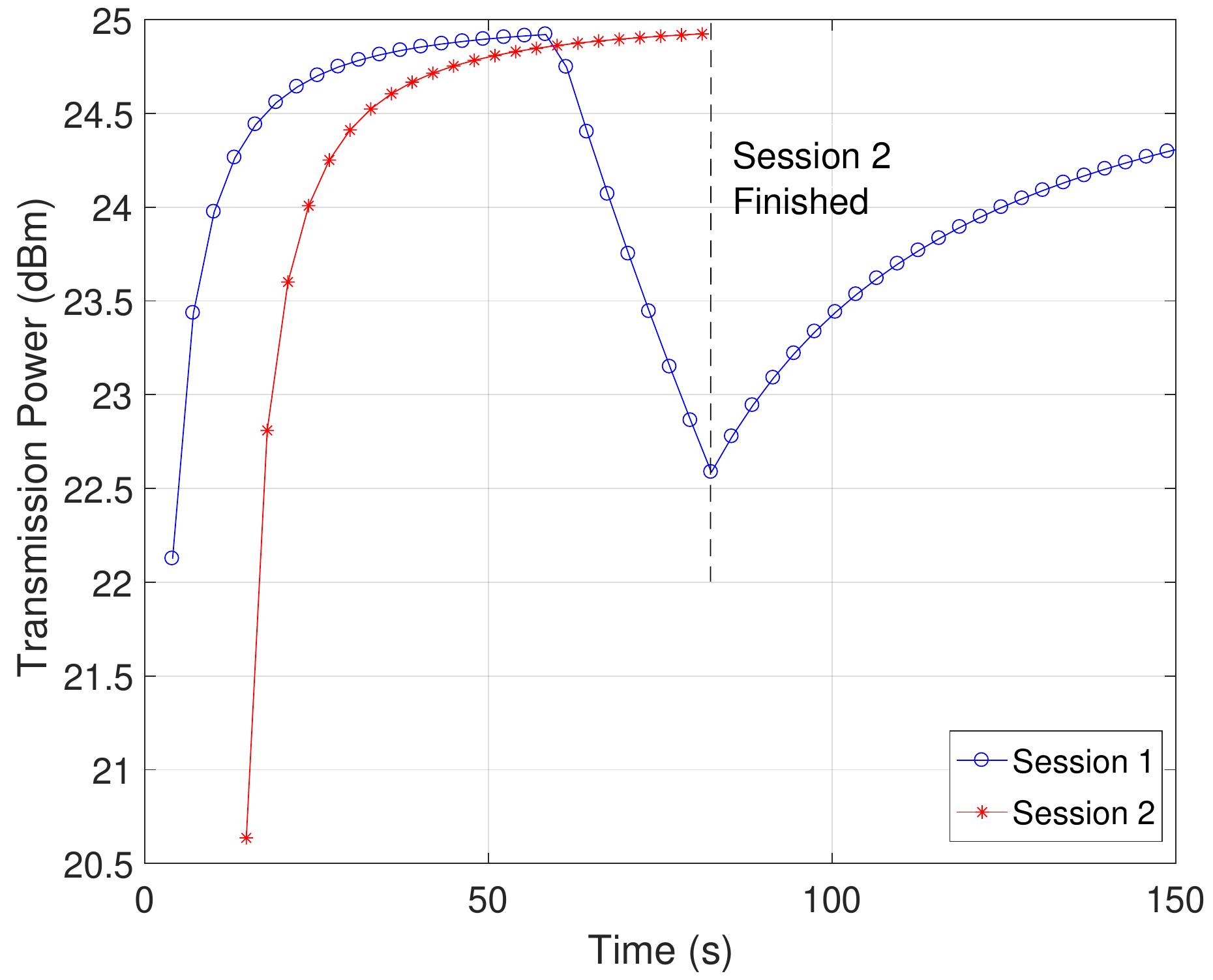}&
\hspace{-3mm}\includegraphics[width=8cm]{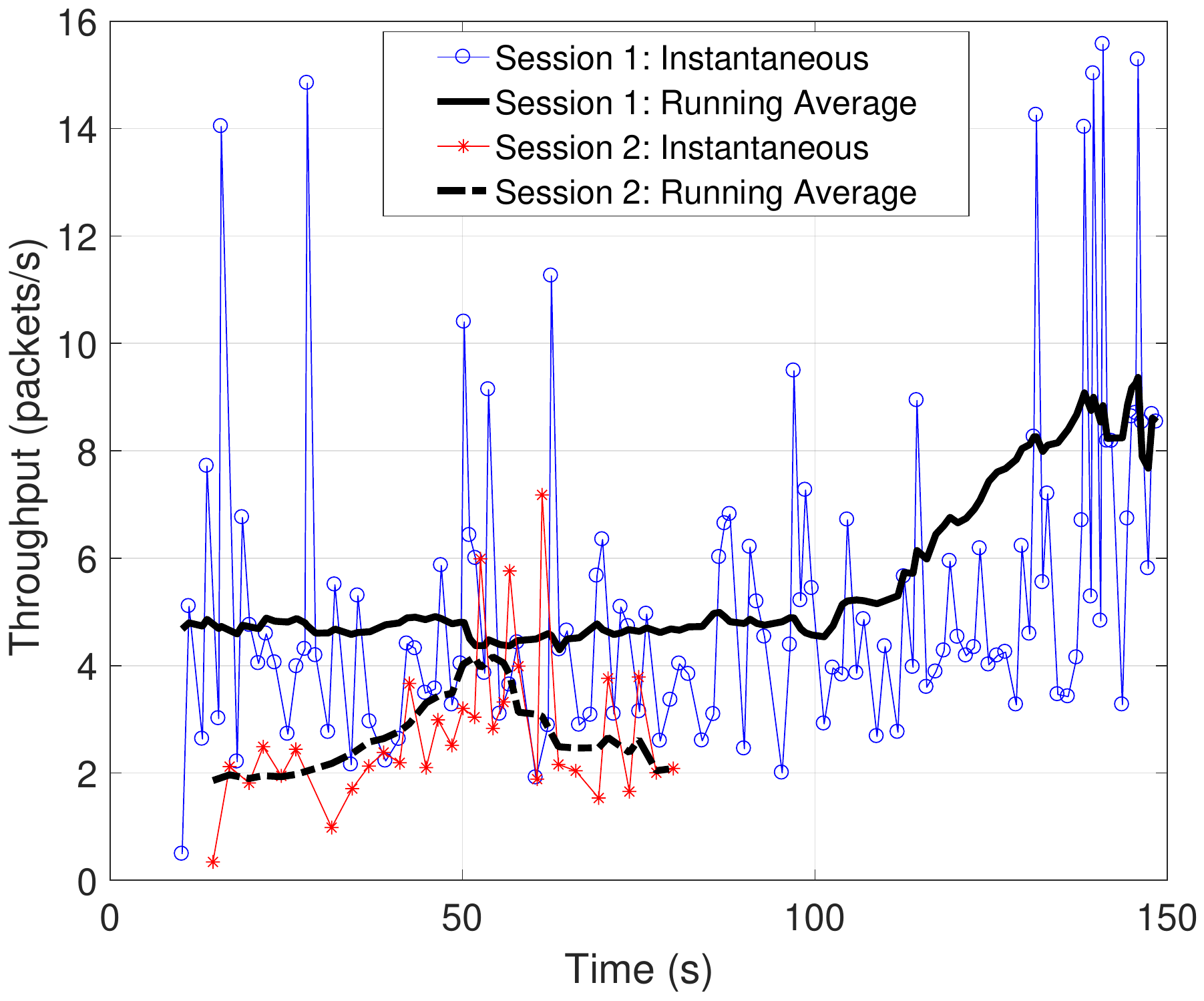}\\
\small (a) & \small (b) \vspace{-4mm}
\end{tabular}
\caption{\small Instance of (a) transmission power (source node) and (b) throughput resulting from sum-log-rate maximization.\vspace{-1mm}} \label{fig:slr}
\end{center}
\end{figure*}

\textbf{Modifying Network Behavior}. In the following experiments, we showcase WNOS's capability of modifying the global network behavior by changing a few lines of code. To achieve different desired network behaviors, one only needs to change the centralized and abstract control objective or modify the constraints while WNOS generates the corresponding distributed control programs automatically. For example, if the control objective is to maximize the sum throughput (i.e., maximize $\sum x $) of all sessions instead of sum log throughput (i.e., maximize $\sum \log(x)$) as in Control Problem 1 (Control Program~2), this can be accomplished by rewriting one line of code only: \emph{expr = mkexpr('sum(wos\_x)', 'wos\_x')}. As shown in Fig.~\ref{fig:twoprob} (top), compared with Control Program~1 (i.e., maximizing sum-log-throughput), Control Program 2 obtains higher sum throughput (4.92 vs 4.66 in $\mathrm{packets/s}$) by increasing the throughput of session 1 while decreasing the throughput of session~2, in this way, as expected, trading throughput for fairness. This is because it is easier for session 1 (see scenario 1 in Fig.~\ref{fig:testtopo}) to achieve higher throughput than session 2 since session 1 has shorter links.

Furthermore, if the network user needs to limit the maximum transmit power of the first session (Control Program 3), this can be accomplished simply by defining a new constraint using the following two lines of code:

\emph{nt.make\_var('wos\_z', [ntses, seslnk, lkpwr], [1, all, None])}

\emph{nt.add\_cstr('wos\_z < 5', 'wos\_z')}\\
where the first line of code defines link power as a variable while the second line specifies the upper bound constraint. The resulting session behaviors are shown in Fig.~\ref{fig:twoprob} (top), where the throughput of session 1 has been effectively bounded. In another example, three sessions are deployed as in scenario 4 in Fig.~\ref{fig:testtopo}. The normalized transmission power of sessions 2 and 3 are programmed to be smaller than 6 and greater than 20, respectively (Control Program 4). It can be seen in Fig.~\ref{fig:twoprob} (bottom) that, compared with Control Program 2, the throughput of sessions 2 and 3 can be successfully changed with the new control program. As shown above, all this needs only two new lines of code to characterize the behavior of session~3.

\textbf{Flexibility and Scalability}. We further test the flexibility and sacalbility of WNOS in changing control programs on a large-scale SDR testbed of 21 USRPs (i.e., Scenario 5) and by considering two sharply different network control objectives: sum-log-rate maximization and power minimization. Again, changing the network control behaviors based on WNOS requires modifying a couple of lines of code only. The \emph{WiNAR} code for defining the power minimization control objective is as follows:

\emph{nt.make\_var(`wos\_x', [ntlk, lkpwr], [all, None]);}

\emph{expr = mkexpr(`sum(wos\_x)', `wos\_x'),}

\noindent where the first line states the transmission power of all the active links in the network as control variables, while the second line defines the \emph{sum of the transmission power} as the utility function to be minimized.



The measured average transmission power of the source and intermediate nodes are plotted in Fig.~\ref{fig:ISECAverage}(a), while the achievable throughput is reported in Fig.~\ref{fig:ISECAverage}(b).
Unsurprisingly, the two control objectives result in different network behaviors. With power minimization, the three sessions achieved approximately the target throughput ($\:\mathrm{packets/s}$) with much lower average power than sum-log-rate maximization; while the latter achieves higher throughput at the cost of higher power consumption.

Figures~\ref{fig:pm} and \ref{fig:slr} provide a closer look at the contrasting network behaviors resulting from the two control objectives, respectively, by plotting the interactions between sessions 1 and 2 in terms of transmission power and the corresponding instantaneous throughputs. It can be seen from Fig.~\ref{fig:pm} that session 2's running average throughput decreases to zero during $20-200\mathrm{s}$ because of low SINR. In response, session 2 increases its transmission power while session 1 decreases until session 2 recovers at around $200\mathrm{s}$. After session 2 is finished, session 1 keeps its current transmission power, which is sufficient to achieve the target throughput. Very differently, in the case of sum-log-rate maximization, after session 2 is done, session 1 increases its transmission power to maximize the throughput, as shown in Fig.~\ref{fig:slr}.

\section{Limitations and Future Work}\label{sec:future}
\vspace{-1mm}
We believe that our work on WNOS provides the first proof of concept of the ability to create a principled optimization-based wireless network operating system, where the desired global network behavior is defined on a centralized high-level
abstraction of the network and obtained through automatically generated distributed cross-layer control programs. We acknowledge several limitations, which will be addressed in future work.

\textbf{Learning-based Automated Modeling}. WNOS generates cross-layer distributed control programs by decomposing high-level defined network control objective problems, and hence users don't have to deal with tedious details of  lower-layer protocols and distributed optimization theory. The decomposition requires the WNOS to specify mathematical models for network protocols at all layers. We are working to standardize the interface of WNOS and  plan to make the source code of WNOS available so that new protocols and mathematical models can be easily incorporated into the existing programmable protocol stack (PPS).
While the current version of the PPS is designed for software defined radios, we also plan to develop versions of the PPS designed to operate on legacy wireless interface cards (e.g., WiFi).
Last, we also plan to extend WNOS to build mathematical models for user-defined network control problems by online learning and automated modeling \cite{Capelo98, Benjamin07}.

\textbf{Multi-timescale Control}. Network protocols at different layers operate at different time scales, which can be up to orders-of-magnitude different. In the current WNOS implementation, static time scales have been considered, e.g., 30 times larger for transport-layer rate adaptation  than physical-layer power control in the testbed evaluation in Section~\ref{sec:evaluation}. In the future, we will work to let WNOS determine time scales automatically for different protocols based on the user-defined high-level network control objective, including convergence and delay requirements, network size, as well as the underlying transmission medium.

\textbf{Decomposition Approaches}. Given user-defined high-level network control problems, mathematical optimization problems are constructed and then decomposed by WNOS. Currently only dual decomposition and decomposition by partial linearization (DPL) have been considered for cross-layer and distributed decompositions, respectively. We plan to incorporate other decomposition approaches, such as primal decomposition, hybrid dual and primal decompositions~\cite{Daniel06}.

\section{Related Work} \label{sec:related}\vspace{-1mm}
This work is related to the notion of software-defined wireless networking, especially for infrastructure-less ad hoc networks.

\textbf{SDN for Infrastructure-based Wireless Networks.} Software-defined networking has shown great potential to enhance the performance of wireless access networks, e.g., improving network resource utilization efficiency, simplifying network management, reducing operating costs, and promoting innovation and evolution \cite{Bansal12, Gudipati13, RcUBe, YapPoster09, Erran12, Akyildiz14, Xavier13, Vanu, Yap09}. Readers are referred to \cite{MaoYang2015, Haque2016} and references therein for excellent surveys of this field.
For example, in \cite{Bansal12} Bansal et al. proposed OpenRadio, which provides a programmable wireless network data plane to enable users/controllers to upgrade and optimize the network in a software-defined fashion.
Demirors et al. proposed RcUBe in \cite{RcUBe}, a new architectural radio framework to provide a programmable protocol stack and to ease the implementation of protocols in a cross-layer fashion.  Gudipati et al. presented SoftRAN~\cite{Gudipati13} to redesign the radio access layer of LTE cellular networks.
In \cite{Erran12}, Li et al. presented CellSDN to simplify the design and management of wireless cellular networks while enabling new applications. 
Different from these works, whose focus is on infrastructure-based cellular networks, here we focus on software-defined wireless networks without an infrastructure (i.e., ad hoc, sensor networks, device-to-device, among others) in which at least a portion of the control process needs to be distributed.

\textbf{SDN for Infrastructure-less Wireless Networks.}  Compared to SDN-based cellular networks, enabling SDN in distributed wireless networks, e.g., multi-hop ad hoc networks and vehicular networks, is significantly more challenging because of the absence of a centralized control entity. Research efforts in this field include \cite{Zhu15, Mehran15, Galluccio15, Alqallaf15, PaoloDiDio16, Zlmmerling12, Miyazaki14, TieLuo2012, DezeZeng17, Vissicchio15}; while \cite{Keshav16, Hassan17} provide a survey of this field. In \cite{Zhu15}, Zhu et al. proposed an SDN-based routing scheme for Vehicular Ad Hoc Network (VANET) where a central controller collects network information from switches and computes the optimal routing strategies. Palazzo et al. proposed SDN-WISE \cite{Galluccio15, PaoloDiDio16} to provide a stateful programmable protocol stack for wireless sensor networks (WSNs).
Zlmmerling et al. proposed pTunes \cite{Zlmmerling12}, a framework for runtime adaptation of low-power MAC protocol parameters in wireless ad hoc networks, where a central base station is used to collect reports on the network state and then determines optimal MAC layer operating point by solving a multi-objective optimization problem. A software-defined wireless sensor network was proposed in \cite{Miyazaki14}, where node behaviors can be redefined at runtime by injecting of sensor node roles via wireless communications. In \cite{TieLuo2012}, Luo et al. proposed Sensor OpenFlow (SOF) as a communication interface between the data and control plane of their designed software-defined WSNs (SD-WSN).

Since most research efforts \cite{Zhu15, Galluccio15, PaoloDiDio16, Zlmmerling12, Miyazaki14, TieLuo2012, DezeZeng17} discussed above rely on a logically-centralized control plane to determine the optimal operating point, the resulting software-defined WSNs suffer in terms of flexibility and scalability because of the significant communication overhead and delay in collecting global network state information. For this reason, in \cite{Vissicchio15} Vissicchio et al. proposed Fibbing to achieve flexibility and robustness through central control over distributed routing in wireless sensor networks. In \cite{Mehran15}, the authors discussed a hybrid SDN architecture for wireless distributed networks to eliminate the need for multi-hop flooding of routing information. In this way, the computational complexity of route discovery is split between the SDN controller and the distributed forwarding nodes, and consequently the SDN controller does not need to collect all the link state information to decide all routes.

Different from existing work, which either relies on an essentially centralized control entity (with limited flexibility and scalability) or focus on a single protocol layer (e.g., the network layer), \emph{we focus on cross-layer control of infrastructure-less wireless networks. Our objective is to study the basic principles for designing WNOS, an optimization-based wireless network operating system}. Based on WNOS, as discussed in Section~\ref{sec:intro}, \emph{distributed cross-layer control programs can be generated automatically based on rigorous distributed optimization theories while the control objectives are defined on a centralized network abstraction provided by WNOS}.

\section{Conclusions} \label{sec:conclusion}
We discussed the basic building principles of the \emph{Wireless Network Operating System (WNOS)}. WNOS provides network designers with an abstraction hiding the lower-level details of the network operations. Based on this abstract representation, WNOS takes centralized network control programs written on a centralized, high-level view of the network and automatically generates distributed cross-layer control programs based on distributed optimization theory that are executed by each individual node on an abstract representation of the radio hardware. We presented the design architecture of WNOS, discussed the technologies to enable automated decomposition of user-defined centralized network control problems. We have also prototyped WNOS and evaluated its effectiveness using testbed results. Future research directions will include automated modeling, multi-timescale control and incorporating heterogeneous decomposition approaches.


\bibliographystyle{ieeetr}

\bibliography{NOS,Mobihoc2010}

\end{document}